\documentclass[a4paper,american,aps,citeautoscript,floatfix,longbibliography,pdftex,pre,
superscriptaddress,showpacs,preprint]{revtex4-1}

\usepackage{adjustbox}

\usepackage{enumerate,tensor}
\usepackage{natbib} 
\usepackage{hyperref}
\usepackage{amsmath,amssymb,amsmath}
\usepackage{graphicx}% Include figure files
\usepackage{dcolumn}% Align table columns on decimal point
\usepackage{bm}% bold math
\usepackage{color}

\newcommand{\ie}{i.\,e.}%

\makeatletter
\def\paragraph{\@startsection{paragraph}{4}{10pt}{-1.25ex plus -1ex minus -.1ex}{0ex plus 0ex}{\normalsize\textit}}
\renewcommand\@biblabel[1]{#1}
\renewcommand\@makefntext[1]%
{\noindent\makebox[0pt][r]{\@thefnmark\,}#1}
\DeclareRobustCommand\onlinecite{\@onlinecite}
\def\@onlinecite#1{\begingroup\let\@cite\NAT@citenum\citealp{#1}\endgroup}
\def\tagform@#1{\maketag@@@{\ignorespaces#1\unskip\@@italiccorr}}
\let\orgtheequation\theequation
\def\theequation{(\orgtheequation)}
\makeatother

\begin{document}

\title{Nonlinear dynamics and energy transfer for two rotating
 dipoles in an external field: A three-dimensional analysis}

\author{Rosario Gonz\'alez-F\'erez}
%\email{rogonzal@ugr.es}%
\affiliation{Instituto Carlos I de F\'{\i}sica Te\'orica y Computacional,
and Departamento de F\'{\i}sica At\'omica, Molecular y Nuclear,
  Universidad de Granada, 18071 Granada, Spain} 

\author{Manuel I\~narrea}
\affiliation{\'Area de F\'{\i}sica, Universidad de La Rioja, 26006 Logro\~no, La Rioja, Spain}

  \author{J. Pablo Salas}
\affiliation{\'Area de F\'{\i}sica, Universidad de La Rioja, 26006 Logro\~no, La Rioja, Spain}

\author{Peter Schmelcher}
\affiliation{The Hamburg Center for Ultrafast Imaging, Luruper Chaussee 149, 22761 Hamburg, Germany}
\affiliation{Zentrum f\"ur Optische Quantentechnologien, Universit\"at
  Hamburg, Luruper Chaussee 149, 22761 Hamburg, Germany} 
  
\date{\today}
\begin{abstract} 
We investigate the structure and the nonlinear dynamics of two rigid polar rotors coupled through the dipole-dipole interaction 
in an  external homogeneous electric field.
In the field-free stable head-tail configuration,
an excess energy is provided to one of the dipoles, and we explore the
resulting three-dimensional classical dynamics.
This dynamics is characterized in terms of the kinetic energy transfer between the dipoles, their orientation
along the electric field, as well as their chaotic behavior.
 The field-free energy transfer mechanism shows an abrupt transition between equipartition
and non-equipartition regimes, which 
 is independent of the initial direction of rotation due to the existence of an infinite set of equivalent manifolds. 
The field-dressed dynamics is highly complex and strongly depends on the electric field strength 
and on the initial conditions.
In the strong field regime, the energy equipartition and chaotic behavior dominate the dynamics.
\end{abstract}
\maketitle

\section{Introduction}
\label{sec:introduction}

The experimental availability of ultracold dipolar gases represents a strong motivation for the
investigation of the physical phenomena  related to the long-range and anisotropic dipole-dipole interaction~\cite{pfau12}.
By tuning the dipole-dipole interaction,  the dipolar gas properties can be significantly modified giving rise to a rich variety of
novel applications, such as,
the creation of  novel quantum liquids~\cite{pfau16,ferlaino16,pfau16_2},  exotic spin 
dynamics~\cite{naylor16,rey19},  and the control of ultracold chemical reactions~\cite{ye10,ye11,ye13}, 
 thermalization~\cite{lev18} or energy exchange~\cite{bakker99,tolbert00,noel15,wenger19}.

The mechanism of energy transport  mediated by the dipole-dipole interaction is closely related to the nonlinear
behavior of  many-body systems and has attracted special attention in  classical 
dynamics~ \cite{michl,sim,jonge,rotors2D,pre2018,estevez18}. 
Most of the nonlinear dynamical studies on coupled rotating dipoles 
restrict their motion to planar rotations. As a consequence, each dipole is described by one angle
and the number of degrees of freedom of the system is equal to the number of dipoles.
A  natural  extension of these studies  is to allow the dipoles to perform three-dimensional (3D) overall rotations,
and, to the best of  our knowledge, the corresponding literature is very scarce. 
The  3D classical dynamics has been explored for isolated rotors, such as diatomic or symmetric top molecules,
 exposed to combinations of external electric fields~\cite{A516,A286}

In our previous work~\cite{rotors2D}, we studied the classical dynamics  
of two polar rotors, coupled by the dipole-dipole interaction, in the presence of an external electric
field in  a planar  invariant manifold.
By restricting the motion of the dipoles to the invariant manifold, the system has two degrees of freedom. The energy transfer mechanism between the dipoles for varying field 
strength has been analyzed in terms of the phase space structure of the system.
Here, we extend this previous study by exploring the complete 3D classical dynamics allowing the dipoles to 
rotate in any direction in space. Since the rotors motion is not restricted to a  planar manifold as in 
Ref.~\cite{rotors2D},  each dipole is described by two angles, and we
encounter a Hamiltonian system with four degrees of freedom. 
We assume that  the dipoles are initially in the stable head-tail configuration with fixed spatial positions.
This stable configuration is perturbed by adding a certain excess energy to one of the dipoles,
which starts rotating from the head-tail configuration axis, while at the same time the electric
field is turned on.
The follow up dynamics is investigated in terms of the energy exchange mechanism and
 the orientations of the rotors induced by the electric field. We also explore the chaoticity of the 
system by using a fast chaos indicator.

Due to the existence of the infinite set of equivalent manifolds, our field-free results show that the energy transfer mechanism is independent of the direction for
which the dipole starts to rotate, which is in complete agreement with  
our previous results~\cite{rotors2D}. 
The field-dressed dynamics is highly complex and strongly depends on the strengths of the 
dipole and the electric field interactions as well as on the initial conditions.
In the very weak field regime, when the dipole interaction is dominant, the
energy transfer mechanism resemble the field-free case, although the border
between the non-equipartition and equipartition regimes becomes more diffuse.
As the electric field increases, the energy equipartition regime dominates the dynamics, 
although we still encounter regions of non-equipartition energy for certain initial conditions.
For strong fields, the two dipoles are significantly oriented along the electric field axis.
When the electric field is turned on, the system becomes non-integrable so that there
appears chaotic motion. For low and intermediate values of the field, the
degree of chaoticity increases as the interaction with the electric field becomes more dominant.
Surprisingly, for strong electric fields the chaoticity of the system remains
very pronounced which is quite unexpected because 
the gradual increase of the field would eventually lead the system to its integrable limit.
Indeed, we verified that, for very large values of the electric field, a slow crossover to an integrable phase space
takes place.

This work is organized as follows. 
In~\autoref{sec:hamiltonian} we discuss the classical Hamiltonian of the system, its
symmetries and  invariant manifolds.  The underlying equilibrium points are also
presented. 
In~\autoref{sec:energytransferfieldfree} we explore the time evolution of the energy transport between 
the dipoles and their orientations with varying   initial conditions  and for several
the electric field strengths. The chaoticity of this systems is discussed in~\autoref{sec:chaos}. 
Our conclusions are provided in~\autoref{sec:con}.
In the Appendix, an exhaustive analysis of the existence, 
stability and bifurcations of the equilibria is given.

\section{Classical Hamiltonian, symmetries and invariant manifolds}
\label{sec:hamiltonian}
\subsection{Classical Hamiltonian}
The potential energy $V_{d}$ between two dipoles with dipole moments ${\bf d}_1$ and
${\bf d}_2$ due to the mutual dipole-dipole interaction (DDI) is
given by~\cite{A753}
\begin{equation}
\label{dipoleInteraction}
V_{d}=\frac{1}{4\pi\epsilon_0}
\frac{({\bf d_1} \cdot {\bf d_2}) \ r^2 - 3 \ ({\bf d_1} \cdot  {\bf r}) \ ({\bf d_2} \cdot  {\bf r})}{r^5},
\end{equation}
with ${\bf r}$ being their relative position.
In our case, we consider two identical rigid rotors
having  electric dipole moments ${\bf d}_1=q \ {\bf l_1}$ and
${\bf d}_2 = q \ {\bf l_2}$, with $d=|{\bf d}_1|=|{\bf d}_2| = q \ l$, being $q$ and $l$ the charge
and length of the dipoles.
\begin{figure}
\includegraphics[width=.7\linewidth]{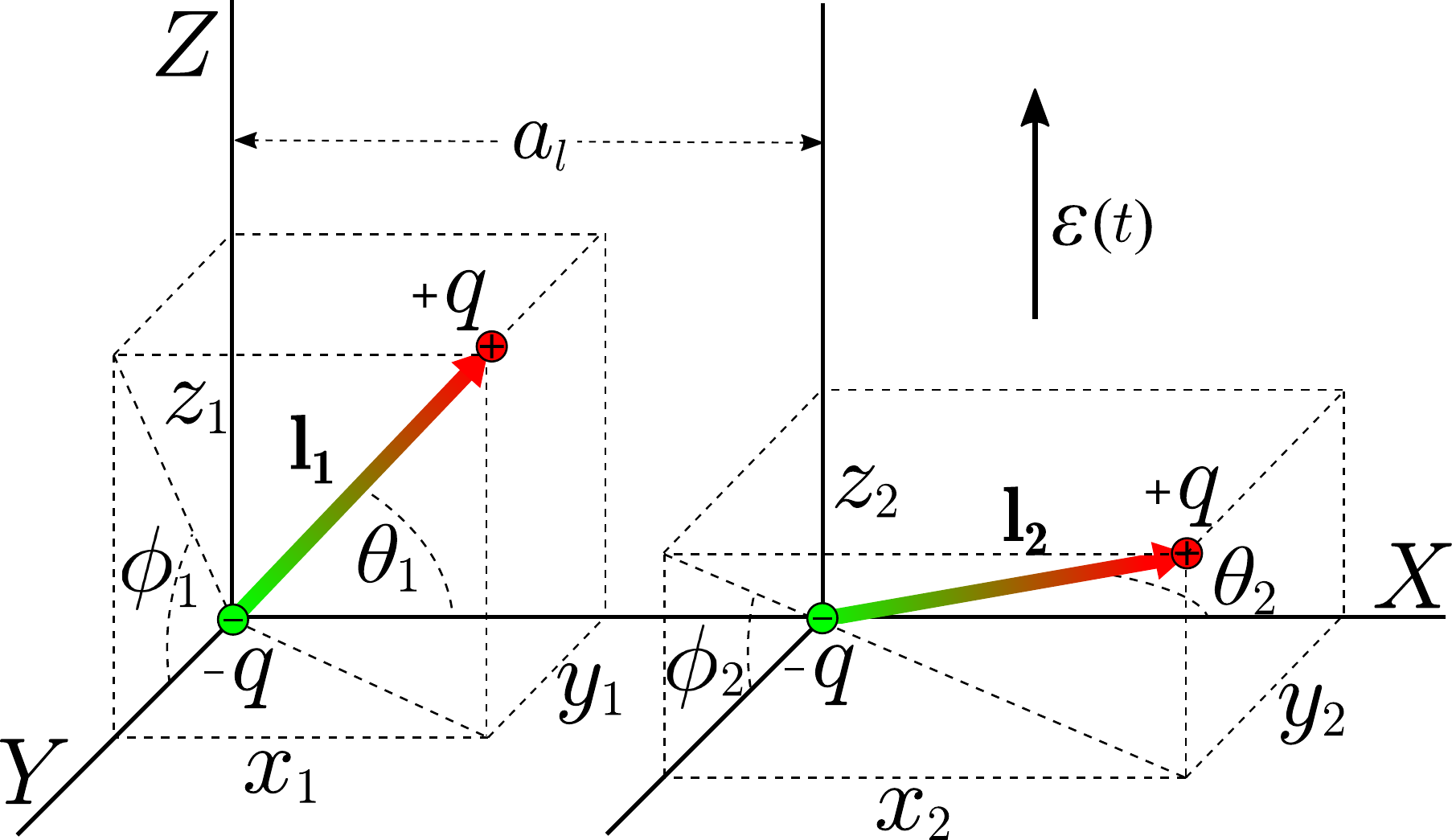}
\caption{Schematic representation of the two interacting dipoles}
\label{fi:dipolos}
\end{figure}
Besides the DDI, the dipoles are in the presence of an external  homogeneous time-dependent electric
field ${\cal E}(t)$ parallel to the  Laboratory Fixed Frame (LFF $XYZ$) $Z$-axis. 
Assuming that the positions of the rotors are fixed along the LFF $X$-axis and separated by a constant 
distance $ a_l$ (see Fig.\ref{fi:dipolos}), 
the total interaction potential
reads as follows
\begin{equation}
\label{potenTotal1}
V(\mathbf{l}_1,\mathbf{l}_2, t)=- q \ {\cal E}(t)(z_1+z_2)+
\frac{q^2}{4\pi\epsilon_0 a_l^3} [\mathbf{l}_1\cdot \mathbf{l}_2 - 3 x_1 x_2],
\end{equation}
where the vectors  $\{\mathbf{l}_i=(x_i, y_i, z_i) || \ , |\mathbf{l}_i|=l, \ i=1, 2 \}$
determine the orientation of the dipoles.
The first term in \ref{potenTotal1} stands for the interaction with the external field and the second one
for the DDI. 
Here, we assume that the homogeneous electric field has a turn-on modeled by
a linear ramp $f(t)$, so that 
${\cal E}(t)= E_s f(t)$ with 
$f(t)$ 
\begin{equation}
\label{pulse}
f(t)=\left \{
      \begin{matrix} 
         \displaystyle  \frac{t}{t_1} & \mbox{if} & 0 \le t < t_1\\[2ex]
         1 & \mbox{if} & t \ge t_1\, 
        \end{matrix}
   \right.
   \end{equation}
This linear ramp was used in Ref.~\cite{chandre} and it also mimics the turn-on of laser pulses
~\cite{A182,A747}.

If $\mu$ is the reduced mass of the dipoles, the dynamics of the system is governed by the classical Hamiltonian
(the energy $E$)
\begin{equation}
\label{hamiCarte}
E\equiv H= \sum_{i=1}^2 \frac{1}{2 \mu}  \bigg[ P_{x_i}^2+
P_{y_i}^2+P_{z_i}^2\bigg]+V(\mathbf{l}_1,\mathbf{l}_2, t),
\end{equation}
subject to 
the holonomic constraints $\{\mathbf{l}_i=(x_i, y_i, z_i) || \ , |\mathbf{l}_i|=l, \ i=1, 2 \}$.
At this point, and without loss of generality, we use a dimensionless version of the
Hamiltonian \ref{hamiCarte} by expressing the energy in units of the parameter
 $\chi=d^2/4\pi\epsilon_0 a_l^3$ that controls the DDI.  To do that, we introduce the dimensionless coordinates
\[
\mathbf{l}'_i=(x'_i, y'_i, z'_i)=(x_i/l, y_i/l, z_i/l), \quad i=1,2,
\]
and the dimensionless time $t'=t/t_d$, where $t_d=\sqrt{I/\chi}$ is the new unit of time
being $I= \mu \ l^2$ the moment of inertia of the dipoles.
After applying these transformations to \ref{hamiCarte}, we arrive at the
following dimensionless Hamiltonian
\begin{equation}
\label{hamiCarte3}
E' \equiv H'= \frac{H}{\chi}= \sum_{i=1}^2 \frac{1}{2}  \bigg[ P_{x_i}'^2+
P_{y_i}'^2+P_{z_i}'^2\bigg]+{\cal V}_1(\mathbf{l}'_1,\mathbf{l}'_2, t').
\end{equation}
where the potential ${\cal V}_1(\mathbf{l}'_1,\mathbf{l}'_2, t')$ reads as
\begin{equation}
\label{potenCarte3}
{\cal V}_1(\mathbf{l}'_1,\mathbf{l}'_2, t')=- \beta f(t') (z'_1+z'_2)+
 (\mathbf{l}'_1\cdot \mathbf{l}'_2 - 3 x'_1 x'_2).
\end{equation} 
In this way, the dynamics depends on the energy $E' \equiv H'=H/\chi$ and
on the new (dimensionless) electric field parameter $\beta=d E_s/\chi$, which is the ratio between the strengths of the electric and the dipole-dipole interactions.
For the sake of simplicity, we omit in the following the primes in
Hamiltonian~\ref{hamiCarte3}

The aforementioned holonomic constraints between the Cartesian coordinates of the dipoles
reduce the number of degrees of freedom
from 6$+1/2$ to 4$+1/2$. 
By taking the LFF $X$-axis as the polar axis,  the transformation between the Cartesian and the Euler angles
$(\theta_1, \phi_1,\theta_2, \phi_2)$ of each rotor reads (see Fig.\ref{fi:dipolos})
\[
x_i=\cos \theta_i,\quad y_i=\sin \theta_i \cos \phi_i,\quad z_i=\sin \theta_i \sin \phi_i, \quad i=1,2
\]
and the Hamiltonian~\ref{hamiCarte3} converts to
\begin{equation}
\label{hamiEuler}
H=\sum_{i=1}^2 \frac{1}{2}  \bigg[ P_{\theta_i}^2+
 \frac{P_{\phi_i}^2}{\sin^2\theta_i}\bigg]+{\cal V}_2(\theta_1, \phi_1, \theta_2, \phi_2; t), 
\end{equation}
where the interaction potential~\ref{potenTotal1} is 
\begin{eqnarray}
\label{potenEuler}
%V &\equiv& 
{\cal V}_2(\theta_1, \phi_1, \theta_2, \phi_2; t)&=&\beta \ f(t)(\sin \theta_1 \sin \phi_1+\sin \theta_2 \sin \phi_2)\nonumber\\[2ex]
 &&+(\sin\theta_1\sin\theta_2 \cos (\phi_{1}-\phi_{2})-2 \cos\theta_1\cos\theta_2).\
 \end{eqnarray}
In spherical coordinates, the Hamiltonian~\ref{hamiEuler}
defines a $(4+1/2)$-degree-of-freedom dynamical system with coordinates $(\theta_1, \phi_1, \theta_2, \phi_2)$ and the
corresponding momenta $(P_{\theta_1},P_{\phi_1}, P_{\theta_2}, P_{\phi_2})$.

\subsection{Symmetries and invariant manifolds}
Since the rotors are identical, Hamiltonians~\ref{hamiCarte3}~and~\ref{hamiEuler}
possess a exchange symmetry.
Besides this discrete symmetry,  the field-free system, \ie, considering only the dipole interaction
and $\beta=0$,  is also invariant under rotations around the 
common LFF $X$-axis. Besides the energy, this continuous symmetry  implies  that  
the $X$-component of the total angular momentum $L_X=P_{\phi_1}+P_{\phi_2}$ is  conserved. 

The Hamiltonian equations of motion arising from~\ref{hamiEuler} read as follows:
\begin{eqnarray}
\label{flux2}
\dot \theta_1 &=&  P_{\theta_1},\, \dot \theta_2 =  P_{\theta_2},
\quad \dot \phi_1 = \frac{P_{\phi_1}}{\sin^2\theta_1},
\, \dot \phi_2 = \frac{P_{\phi_2}}{\sin^2\theta_2}\nonumber\\[2ex]
\dot P_{\theta_1} &=& \frac{P^2_{\phi_1} \ \cos\theta_1}{\sin^3\theta_1}+\beta f(t) \cos \theta_1 \sin \phi_1
\nonumber\\[2ex]
&-&\cos\theta_1\sin\theta_2 \cos\phi_{12} -2 \sin\theta_1\cos\theta_2,\nonumber \\[2ex]
\dot P_{\theta_2} &=& \frac{P_{\phi_2}^2 \cos\theta_2}{\sin^3\theta_2}+\beta f(t) \cos \theta_2 \sin \phi_2
\nonumber\\[2ex]
&-&\sin\theta_1\cos\theta_2 \cos\phi_{12} -2 \cos\theta_1\sin\theta_2,\\[3ex]
\dot P_{\phi_1} &=& \beta f(t) \sin \theta_1 \cos \phi_1+
\sin\theta_1\sin\theta_2 \sin\phi_{12}.\nonumber\\[3ex]
\dot P_{\phi_2} &=& \beta f(t) \sin \theta_2 \cos \phi_2
-\sin\theta_1\sin\theta_2 \sin\phi_{12}.\nonumber
\end{eqnarray}
\noindent
For $\beta=0$, the infinite set of manifolds ${\cal M}$ of codimension
four given by
\begin{eqnarray}
\label{manifoldM}
{\cal M} = \lbrace(\theta_1, P_{\theta_1}, \theta_2, P_{\theta_2}) \ | \ \phi_1-\phi_2=0,\nonumber\\
 \ \mbox{and} \ P_{\phi_1}=P_{\phi_2}=0; \ \beta=0\rbrace.
\end{eqnarray}
are invariant under the dynamics. 
On each of these manifolds, the Hamiltonian~\ref{hamiEuler} for $\beta=0$
reduces to the two degrees of freedom Hamiltonian
\begin{equation}
\label{hamiEuler4}
{\cal H}_{\cal M} =\frac{P^2_{\theta_1}+P^2_{\theta_2}}{2}+
 \sin\theta_1\sin\theta_2 -2 \cos\theta_1\cos\theta_2,
\end{equation}
\noindent
and the rotational motion of the dipoles
is restricted to a given common polar plane of constant azimuthal inclination 
$\phi_1-\phi_2=0$ where the polar angles $(\theta_1, \theta_2)$ vary in the interval $[-\pi, \pi)$.
The existence of the manifolds ${\cal M}$ is associated to the aforementioned rotational invariance
of the Hamiltonian~\ref{hamiCarte3}
around the LFF $X$-axis.
For $\beta \ne 0$, the electric
field breaks this rotational symmetry, and $L_X=P_{\phi_1}+P_{\phi_2}$
is no longer an integral of the motion. It is worth noticing that the presence of the electric field
reduces the (infinite) invariant manifolds ${\cal M}$ to a single one along the direction
$\phi_1=\phi_2=0$, namely
\begin{eqnarray}
\label{manifoldMxz}
{\cal M}_{XZ} = \lbrace(\theta_1, P_{\theta_1}, \theta_2, P_{\theta_2}) \ | \ \phi_1=\phi_2=0,\nonumber \\
\ \mbox{and} \ P_{\phi_1}=P_{\phi_2}=0\rbrace.
\end{eqnarray}
The planar dynamics and the energy transfer on the manifold ${\cal M}_{XZ}$ has already
been studied in Ref.~\cite{rotors2D}.

The equations of motion~\ref{flux2} provide
an additional invariant manifold ${\cal N}$ of codimension four,
\begin{equation*}
\label{manifoldN}
{\cal N} = \lbrace(\phi_1, P_{\phi_1}, \phi_2, P_{\phi_2}) \ | \ \theta_1=\theta_2=\pi/2, \ P_{\theta_1}=P_{\theta_2}=0\rbrace.
\end{equation*}
When the system is moving on this manifold ${\cal N}$, the dynamics is
governed by the $(2+1/2)$-degree-of-freedom Hamiltonian
\begin{equation*}
%\label{hamiEuler5}
{\cal H}_{\cal N}=\frac{P_{\phi_1}^2}{2}+\frac{P_{\phi_2}^2}{2}-\beta f(t) (\sin \phi_1+
\sin \phi_2)+\cos\phi_{12}.
\end{equation*}
On the manifold ${\cal N}$, the rotational motion of the dipoles is restricted to the parallel $y_1-z_1$ and
$y_2-z_2$ planes, respectively. This configuration was already considered in Ref.~\cite{jonge}
for the general case of a chain of $N\ge2$ dipoles and zero electric field.

\subsection{Equilibrium configurations}
\label{sec:equilibria}
The equilibrium points of a dynamical system provide useful information about its behavior.
For $t \ge t_1$,  the electric field parameter reaches its maximal value $\beta$ with $f(t)=1$, and,
 using the Cartesian  %formulation given by
Hamiltonian~\ref{hamiCarte3}, the equilibrium points  are the critical points of the potential
\begin{equation}
\label{potencarte}
{\cal V}_1(\mathbf{l}_1, \mathbf{l}_2)=- \beta (z_1+z_2) +[\mathbf{l}_1\cdot \mathbf{l}_2 - 3 x_1 x_2]
 % (y_1 y_2 + z_1 z_2 - 2 x_1 x_2),
%\qquad t \ge t_1,
\end{equation}
which is ${\cal V}_1(\mathbf{l}_1,\mathbf{l}_2, t)$~\eqref{potenCarte3} with   $t \ge t_1$, 
under the constraints $\{ \mathbf{l}_i= (x_i, y_i, z_i) || \, |\mathbf{l}_i|^2=1, \ i= 1, 2\}$, 
%$\{(x_i, y_i, z_i) || \ x_i^2+y_i^2+z_i^2=1, \ i= 1, 2\}$, 
together with 
the conditions $P_{x_i}=P_{y_i}=P_{z_i}=0$.
Thence, by introducing the
Lagrange multipliers $\lambda_1$ and $\lambda_2$, the critical points of~\ref{potencarte} are the extrema of 
the Lagrange function ${\cal V}_L$,
\begin{eqnarray}
\label{lagrange}
{\cal V}_L(\mathbf{l}_1, \mathbf{l}_2, \lambda_1, \lambda_2)&=& {\cal V}_1(\mathbf{l}_1, \mathbf{l}_2)%- \beta (z_1+z_2) + (y_1 y_2 + z_1 z_2 - 2 x_1 x_2) +
 %\lambda_1 (1-x_1^2-y_1^2-z_1^2)
\\[2ex]
 &+&
\lambda_1(1-|\mathbf{l}_1|^2)+\lambda_2(1-|\mathbf{l}_2|^2)  \nonumber
%\lambda_2 (1-x_2^2-y_2^2-z_2^2).
\end{eqnarray}
Thus, the critical points are roots of the system of equations
$\vec\nabla_{x_i, y_i, z_i, \lambda_1, \lambda_2} {\cal V}_L=0$, given by 
\begin{eqnarray}
&&x_2+\lambda_1 x_1=0,\quad x_1+\lambda_2 x_2=0, \nonumber\\[2ex]
\label{fluxyz}
&&y_2-2\lambda_1 y_1=0,\quad y_1-2\lambda_2 y_2 =0,\\[2ex]
&&z_2-2\lambda_1 z_1-\beta=0,\quad z_1 -2\lambda_2 z_2-\beta=0 \nonumber,\\[2ex]
&& 1-|\mathbf{l}_1|^2=0,\quad 1- |\mathbf{l}_2|^2=0 \nonumber.
%&& 1-x_1^2-y_1^2-z_1^2=0,\quad 1-x_2^2-y_2^2-z_2^2=0 \nonumber.
\end{eqnarray}
In the presence of an electric field ($\beta \ne 0$), the roots of~\ref{fluxyz} provide the
following six sets of (isolated) critical points:
\begin{itemize}
\item[i)] Two critical points $P_1$: $z_1=z_2=\beta/3$, $y_1=y_2=0$,
$x_1=x_2=\pm \sqrt{1-\beta^2/9}$.
These points exist when $\beta < 3$ and their
 energy is $E_1=-(6+\beta^2)/3$. In the field-free case, \ie, for $\beta=0$, they correspond to the well-known stable {\sl head-tail} configurations of the two dipoles along the common LFF $X$-axis.

\item[ii)] Two critical points $P_2$: $z_1=-z_2=\pm 1$ and $x_1=x_2=y_1=y_2=0$.
The energy of these points is $E_2=-1$ and they exist for values $\beta>0$.

\item[iii)] The critical point $P_3$: $z_1=z_2=1$ and $x_1=x_2=y_1=y_2=0$.
Its energy is $E_3=1-2\beta$, and it exists for $\beta>0$.

%When $\beta \le 3\chi$, $P_3$
%is stable, while $\beta > 3\chi$, it changes its stability becoming a saddle point.

\item[iv)] The critical point $P_4$:  $z_1=z_2=-1$ and $x_1=x_2=y_1=y_2=0$.
The energy of this point is $E_4=1+2\beta$ and it exists for $\beta>0$.

%When $\beta \le \chi$, $P_4$
%is a saddle point, while $\beta > \chi$, it becomes a maximum.

\item[v)] Two critical points $P_5$: $z_1=z_2=-\beta$, $y_1=y_2=0$, 
$x_1=-x_2=\pm \sqrt{1-\beta^2}$. These points exist when $0\le \beta < 1$ and their
 energy is $E_5=(2+\beta^2)$. In the absence of an
electric field $\beta=0$, they correspond to the well-known
unstable {\sl tail-tail} or {\sl head-head} configurations
of the two dipoles along the common LFF $X$-axis.

\item[vi)] Two critical points $P_6$:  $z_1=z_2=\beta/2$, $y_1=-y_2=\pm \sqrt{1-\beta^2/4}$,
$x_1=x_2=0$. These points exist when $0\le \beta < 2$ and their
 energy is $E_6=-(2+\beta^2)/2$.
\end{itemize}
In contrast, for the field-free case $\beta=0$, the number of critical points reduces to $P_1$
and $P_5$,  and to two degenerate circles of equilibria $D_1$ and $D_2$:
\begin{itemize}
\item The set $D_1$ is given by $z_1=z_2=\cos\alpha$, $y_1=y_2=\sin\alpha$,
$x_1=x_2=0$, with $\alpha=[0, 2\pi)$. The energy of this circle of
stationary points is $E_3=1$,
and, when $\beta=0$, the former isolated critical points
$P_3$ and $P_4$ for $\beta \ne 0$ are included in $D_1$.
\item The set $D_2$ is given by $z_1=-z_2=\cos\alpha$, $y_1=-y_2=\sin\alpha$,
$x_1=x_2=0$, with $\alpha=[0, 2\pi)$. The energy of these equilibria is $E_4=-1$,
and, when $\beta=0$, the former isolated equilibria $P_2$ and $P_6$ for $\beta \ne 0$ are included in $D_2$.
\end{itemize}
A detailed study of the stability and existence
of the critical points as well as their bifurcations  as $\beta$ is varied is provided in the Appendix.

Let us explain the differences between the equilibria of the full 3D system and 
when the dynamics is restricted to the planar manifold ${\cal M}_{XZ}$~\cite{rotors2D}.
%At this point, we emphasize the different results encountered when the equilibria
%are studied in the full 3D system with respect to the results when the equilibria are studied
%in the planar manifold ${\cal M}_{XZ}$ (see \cite{rotors2D}).
For $\beta = 0$, there are  two isolated equilibria $P_{1,5}$ and the degenerate circles of equilibria $D_{1,2}$; whereas 
on the manifold ${\cal M}_{XZ}$, there are five isolated equilibria $P_{1,2,3,4,5}$.
For  $\beta \ne 0$, all the equilibria are isolated, such that in the 3D system there appears an additional equilibrium
$P_6$ located off the manifold ${\cal M}_{XZ}$.

\section{Energy transfer processes}
\label{sec:energytransferfieldfree}
In this section,  we explore the classical dynamics of the two dipoles in an external electric field. 
We assume that initially at $t=0$, the two dipoles are at rest, with zero kinetic energy,
 in the stable {\sl head-tail} configuration along the LFF $X$-axis
 given by the equilibrium points $P_1$. 
In this state, the system has the minimal energy $E_1=-2$.
From this initial configuration, the field is turned on at $t=0$ by the ramp-up 
function~\ref{pulse}, and a certain excess energy $\delta K$ is given to dipole one. Therefore, taking into account 
the holonomic constraints $\{\mathbf{l}_i=(x_i, y_i, z_i) || \ |\mathbf{l}_i|^2=1, i=1, 2 \}$, the system leaves the stable equilibrium configuration in such a way that, at $t=0$,  the initial conditions of the dipoles are
\begin{eqnarray}
\label{ci1}
x_1(0)&=&x_2(0)=1, \quad y_1(0)=y_2(0)= z_1(0)=z_2(0)=0\nonumber\\[2ex]
P_{y_1}(0)& =&\sqrt{2\delta K} \ \cos \alpha,  \quad P_{z_1}(0)=\sqrt{2\delta K} \  \sin \alpha, 
 \nonumber
\\[2ex]
P_{x_1}(0)&=&P_{x_2}(0)=P_{y_2}(0)=P_{z_2}(0)=0.\label{ciP1}
\end{eqnarray}
The angle $\alpha \in [-\pi,\pi)$  allows us  to consider all possible
directions of rotations perpendicular to the head-tail axis, and
% in which the dipole one  begins to rotate, and, therefore, 
all  initial conditions of its  momentum. 
Using the initial conditions~\ref{ci1}, the equations of motion given by the Hamiltonian~\ref{hamiCarte3} are numerically integrated up to 
a final time $t_f$ by means of the so-called St\"ormer-Verlet algorithm~\cite{Andersen}. This numerical algorithm is a symplectic 
integrator that preserves the holonomic constraints of the system. The final integration time is fixed to $t_f=400$. 
Our numerical tests have shown that this stopping or final time is appropriate for  a proper characterization of the outcomes.
 In  particular, for a LiCs molecule in its ground state trapped in an
 optical lattice with $a_l=450$ nm, the time unit is
 $t_d=\sqrt{I/\chi} \approx 6.5$~ns, and the final time $t_f=400 \approx 2500$~ns. 
The ramp-up time is fixed to $t_1 = 2$, that roughly corresponds to $12$~ns and that can be achieved in current experiments with realistic field strengths.

For these initial conditions, we explore the dynamics of this system for different ratios of the electric field interaction and the dipole-dipole one,
specifically $\beta=0, 0.1, 1, 10, 100, 1000$. 
To do so, we compute the normalized time-average of the kinetic energy of each dipole, $\widehat K_i$,  given by
\begin{equation}
\label{average}
\widehat K_i=\frac{\langle K_i \rangle}{\langle K_1 \rangle + \langle K_2 \rangle},
\end{equation}
with
\begin{equation}
\label{average}
\langle K_i \rangle = \frac{1}{2 \, t_f} \int_{0}^{t_f} [P_{xi}^2(t)+P_{yi}^2(t)+P_{zi}^2(t)] dt,
\end{equation}
where $t_f$ is the final time. In the presence of the electric field, the axial symmetry no 
longer exists  and the dipoles tend to orient along the  electric field direction. We characterize 
their orientations by computing the time-average of the Cartesian coordinate $z_i$ 
\begin{equation} 
\langle z_i \rangle = \frac{1}{t_f} \int_{0}^{t_f} z_{i}(t) dt, \qquad i=1,2.
\end{equation} 
%Thence, we investigate the energy transfer between the dipoles and their orientation as a function of 
%the kinetic energy $\delta K$ given to dipole one  and of  the direction in which this dipole begins to rotate, \ie, the angle $\alpha$.  
Computationally, these time-averaged quantities $\widehat K_i$ and $\langle z_i \rangle$ are easily 
calculated as a function of  the kinetic energy $\delta K$ given to dipole one  and of  the angle 
$\alpha$ with which this dipole begins to rotate. 
%for different values of the angle $\alpha$. 
However, from a realistic perspective, this angle is difficult to be externally 
controlled. In this way, we also compute  the averages of $\widehat K_i$ and $\langle z_i \rangle$ over all possible realizations 
of this  angle $\alpha$.
These $\alpha$-averaged quantities $\langle \widehat K_i \rangle_{\alpha}$ and 
$\langle z_i \rangle_{\alpha}$   provide insight, as a function of the excess energy $\delta K$, 
into the global extent of the energy equipartition 
and the orientation of the dipoles, respectively.

\begin{figure}
\includegraphics[width=.99\linewidth]{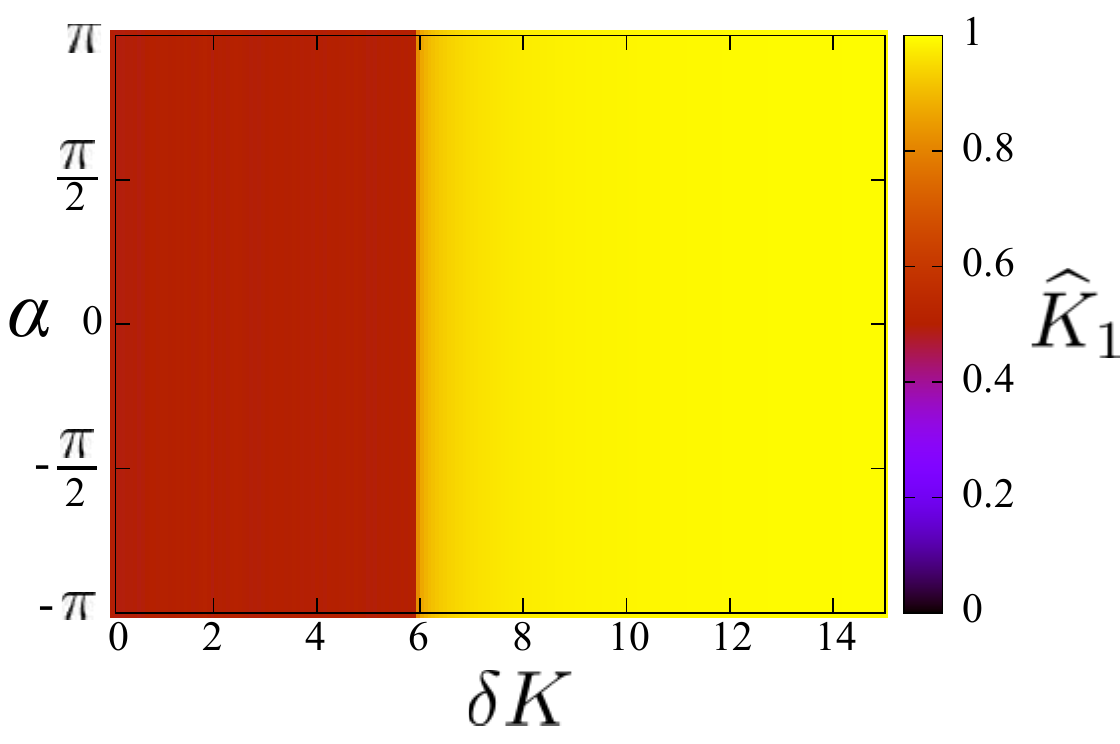}
\caption{Normalized time-averaged kinetic energy $\widehat K_1$ of the first dipole, for $\delta K \in [0.1, 15]$ and $\alpha \in [-\pi,\pi)$, in the field-free case $\beta=0$.}
\label{fi:Mapa1E0}
\end{figure}
Let us start analyzing the  field-free system, \ie, $\beta=0$. The dynamics depends on  $\delta K$, which 
is the excess kinetic energy given to the system, and  on the angle $\alpha$. The normalized time-averaged  kinetic 
energy of the dipole one is presented in~\autoref{fi:Mapa1E0}  for $0 \le \delta K \le 15$ and 
$\alpha \in [-\pi,\pi)$. According to this color map, the kinetic energy transfer between the dipoles 
does not depend on $\alpha$, that is, on the direction in which the first dipole begins to rotate. In the 
initial state  both dipoles are aligned along the symmetry axis of the field-free system, the LFF $X$-axis, 
which is the generatrix of the invariant manifolds ${\cal M}$. Since the excess kinetic energy is added 
to one of the dipoles, the system will be always moving on one of these invariant manifolds ${\cal M}$. 
As the dynamics on these manifolds ${\cal M}$ is equivalent, due to the rotational symmetry 
around the LFF $X$-axis, the direction in which dipole one starts to rotate has no impact on the global 
dynamics. Thus, the normalized time-averaged kinetic energy %with $\delta K$
is independent of  $\alpha$, as it is shown in~\autoref{fi:Mapa1E0}.

For an excess energy $\delta K$ smaller than the  critical value 
$\delta K_c \approx 6$,~\autoref{fi:Mapa1E0} shows that the system always reaches the equipartition 
energy regime,  $\langle K_1 \rangle$ is very close to $\langle K_2 \rangle$,  and a continuous 
energy flow between the rotors occurs. For $\delta K\approx 6$, this equipartition regime abruptly 
breaks, so that most of the kinetic energy remains always in dipole one for $\delta K\gtrsim 6$. As a 
consequence,  the equipartition energy regime inside the invariant manifolds ${\cal M}$ applies only 
for low values of the excess energy $\delta K<\delta K_c$. This feature was detected in Ref.~\cite{jonge}. The authors provided a dynamical explanation of this phenomenon 
in~\cite{rotors2D} when they studied the energy transfer only in the invariant manifold  ${\cal M}_{XZ}$ 
given by~\ref{manifoldMxz}. Let us emphasize that taking into account all the invariant manifolds 
${\cal M}$,~\autoref{fi:Mapa1E0} proves that the system exhibits the same feature in all of them, as they 
are dynamically equivalent due to the field-free axial symmetry.

\begin{figure}
\includegraphics[width=.99\linewidth]{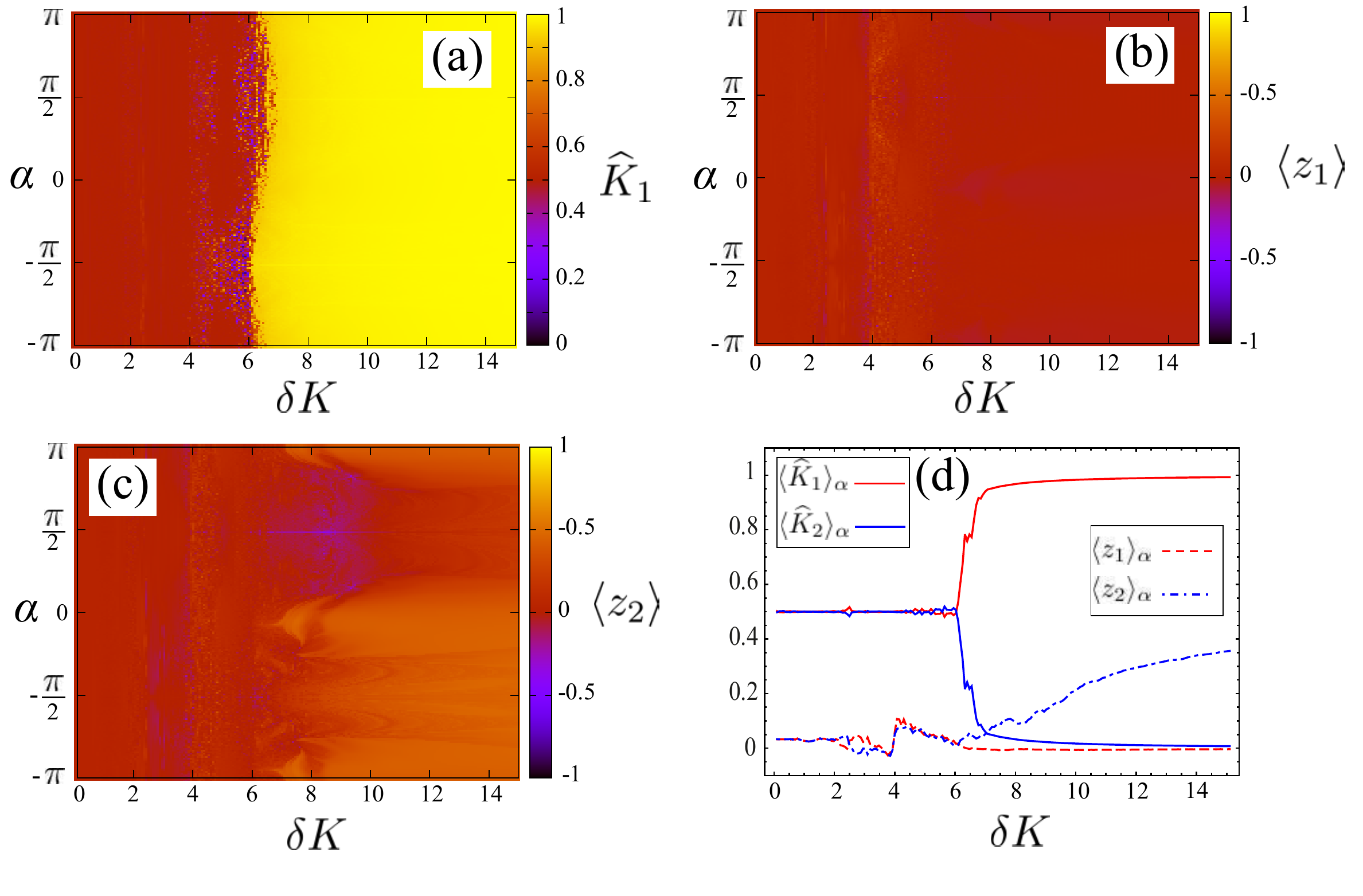}
\caption{(a) Normalized time-averaged kinetic energy $\widehat K_1$ of the first dipole. (b)-(c) Time
averages $\langle z_1 \rangle$ and $\langle z_2 \rangle$, respectively. (d) The
$\alpha$-averaged $\langle \widehat K_i \rangle_{\alpha}$ and $\langle z_i \rangle_{\alpha}$
over the internal distribution angle $\alpha$. For all  panels, the  electric field parameter is $\beta=0.1$.}
\label{fi:Mapa2E01}
\end{figure}

For a weak electric field, the electric field interaction can be considered as a perturbation to the 
dipole-dipole one. Indeed,  for $\beta=0.1$, the normalized time-averaged kinetic energy of
dipole one $\widehat K_1$  in~\autoref{fi:Mapa2E01}(a) shows qualitatively a similar behaviour 
as the field-free results  of~\autoref{fi:Mapa1E0}, except for a region around the critical value 
$\delta K_c=6$. Thence, for $\delta K\lesssim 6$,  the system reaches the equipartition regime, 
while for $\delta K\gtrsim 6$ most of the kinetic energy remains stored in dipole one. It is worth 
noticing that, in the neighborhood of $\delta K_c\approx6$, the time-averaged $\widehat K_1$ shows 
sudden (irregular) variations, which indicate that in this region the system is very sensitive to the 
initial conditions,  \ie, to the values of $\delta K$ and $\alpha$. The evolution of $\langle z_1 \rangle$ 
depicted in~\autoref{fi:Mapa2E01}(b) indicates that this weak electric 
field  $\beta=0.1$ does not cause any significant orientation in dipole one.
In contrast, for  $\delta K\gtrsim 6$, the color map of~\autoref{fi:Mapa2E01}(c) shows
regions of initial conditions leading to a moderate orientation of the dipole two along the electric field
with $\langle z_2 \rangle \approx 0.5$, see the light brown regions of~\autoref{fi:Mapa2E01}(c).
A correlation between the behaviors of $\widehat K_i$ and $\langle z_i \rangle$ arises when the 
$\alpha$-averaged  $\langle \widehat K_i \rangle_{\alpha}$ and $\langle z_i \rangle_{\alpha}$ are 
analyzed in~\autoref{fi:Mapa2E01}(d). While the system is in the
equipartition region, \ie, $\delta K\lesssim 6$, the orientation of both dipoles
is negligible $\langle z_i \rangle_{\alpha}\approx 0$. 
However, in the region $\delta K\gtrsim 6$ where most of the kinetic energy is
in dipole one, the $\alpha$-averaged orientation $\langle z_2 \rangle_{\alpha}$ of dipole two begins to
increase monotonically with $\delta K$, whereas the dipole one remains
non-oriented $\langle z_1 \rangle_{\alpha}\approx 0$.
This behavior is expected  because in the non-equipartition regime 
most of the kinetic energy is stored in dipole one,  which swings very fast
compared to dipole two. Thence, dipole two is more likely to be oriented along  the field  than dipole one.

\begin{figure}[t]
\includegraphics[width=.99\linewidth]{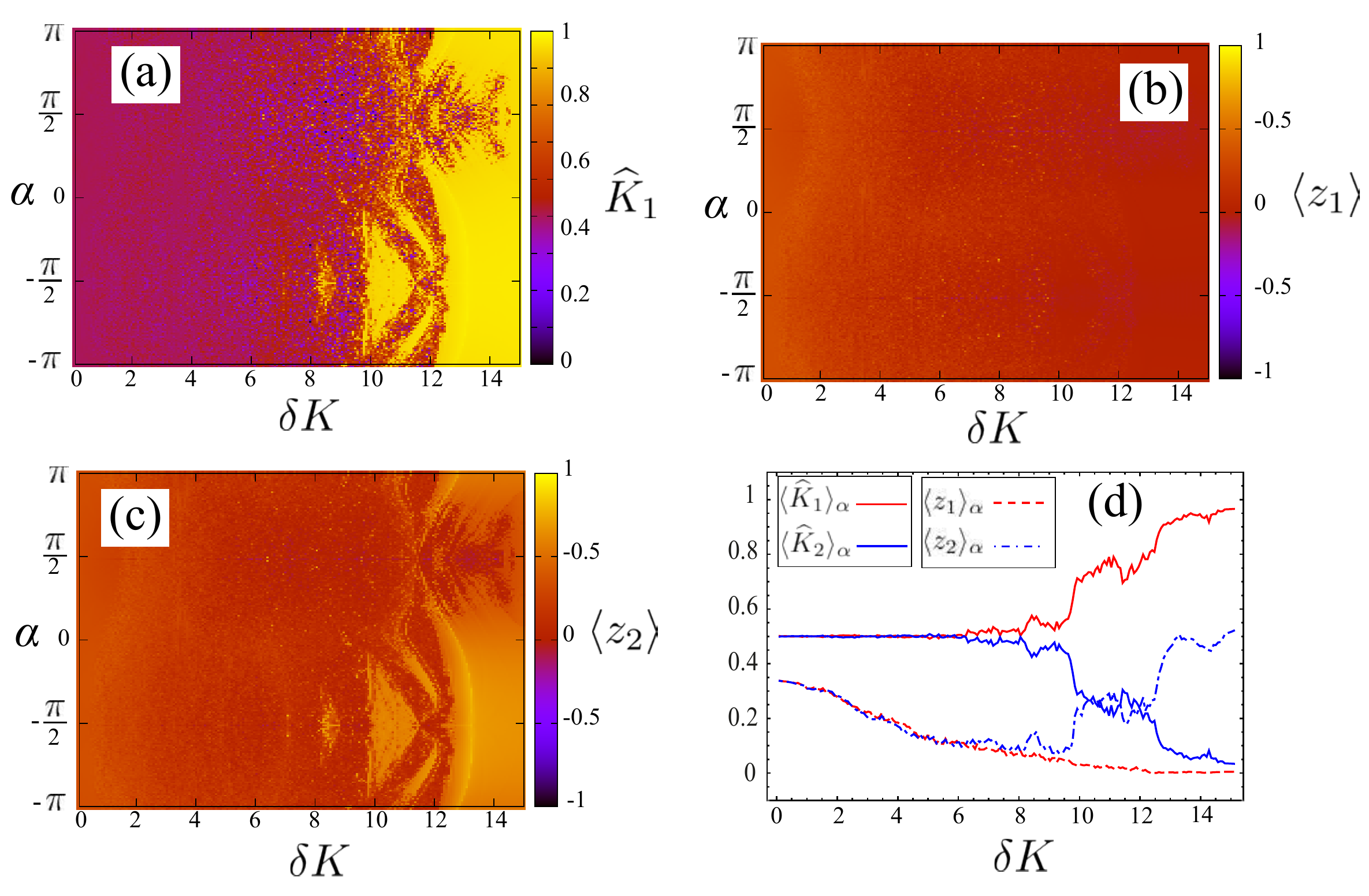}
\caption{(a) Normalized time-averaged kinetic energy $\widehat K_1$ of the first dipole. (b)-(c) Time
averages $\langle z_1 \rangle$ and $\langle z_2 \rangle$, respectively. (d) The
$\alpha$-averaged $\langle \widehat K_i \rangle_{\alpha}$ and $\langle z_i \rangle_{\alpha}$
over the internal distribution angle $\alpha$. 
All the panels for an electric field parameter $\beta=1$.}
\label{fi:Mapa3E1}
\end{figure}

Now, we analyze the dynamics when the dipole-dipole and the electric field interactions are of 
the same order of magnitude, specifically we take  $\beta=1$.
The corresponding results are shown in~\autoref{fi:Mapa3E1}.
Roughly speaking, for $\delta K\lesssim 8$ the system eventually relaxes to equipartition, 
see~\autoref{fi:Mapa3E1}(a). Whereas, for $\delta K\gtrsim 8$ the system progressively moves
away from equipartition, and as  $\delta K$ increases, most of the kinetic energy remains in the
initially excited dipole one.  For $\delta K\gtrsim 8$, it is worth noticing  the remarkable complex behavior
observed in the evolution of $\widehat K_1$  in~\autoref{fi:Mapa3E1}(a), \ie, the dynamics  
shows a high sensitivity to  the initial conditions $\delta K$ and $\alpha$. 
This feature will be addressed in the next section.
For $\delta K\lesssim 8$, there is no remarkable difference in  the orientations of the dipoles 
in~\autoref{fi:Mapa3E1}(b)-(c). We only highlight the slight orientation of both dipoles for 
small values of $\delta K$, see the light brown regions in Figs.~\ref{fi:Mapa3E1}(b)-(c) for 
$\delta K\lesssim 2$. By further increasing the excess energy, $\delta K\gtrsim 8$, the
orientation of each dipole evolves differently. While the orientation of dipole one slightly 
decreases for increasing $\delta K$, see~\autoref{fi:Mapa3E1}(b),  there are regions 
in~\autoref{fi:Mapa3E1}(c) where the orientation of the dipole two presents a significant 
increase. Furthermore, Figs.~\ref{fi:Mapa3E1}(a) and (c) show similar patterns, which
confirm a clear correlation between the non-equipartition regime with most of the kinetic
energy in dipole one, and dipole two having a larger orientation.

The evolutions of the $\alpha$-averaged $\langle \widehat K_1 \rangle_{\alpha}$ and
$\langle z_i \rangle_{\alpha}$ depicted in~\autoref{fi:Mapa3E1}(d) show that, indeed, 
there is equipartition up to $\delta K \lesssim 8$. For this situation, 
$\langle z_1\rangle_{\alpha}$ and $\langle z_2 \rangle_{\alpha}$ are equal and take the
maximal orientation for small excess energy values, and they have very similar values 
and decrease monotonically for increasing $\delta K$. When the equipartition regime is 
lost for $\delta K\gtrsim 8$, the evolutions  of the $\alpha$-averaged  
$\langle \widehat K_1 \rangle_{\alpha}$ and $\langle \widehat K_2 \rangle_{\alpha}$, 
red and blue solid lines in~\autoref{fi:Mapa3E1}(d), respectively, indicate
a gradual but non-monotonic growth of the kinetic energy of dipole one at expense of 
the one of dipole two. At the same time, $\langle z_1 \rangle_{\alpha}$ continues decreasing 
for increasing $\delta K$, whereas, analogously to $\langle \widehat K_1 \rangle_{\alpha}$, 
$\langle z_2 \rangle_{\alpha}$ non-monotonically increases.

\begin{figure}[t]
\includegraphics[width=.99\linewidth]{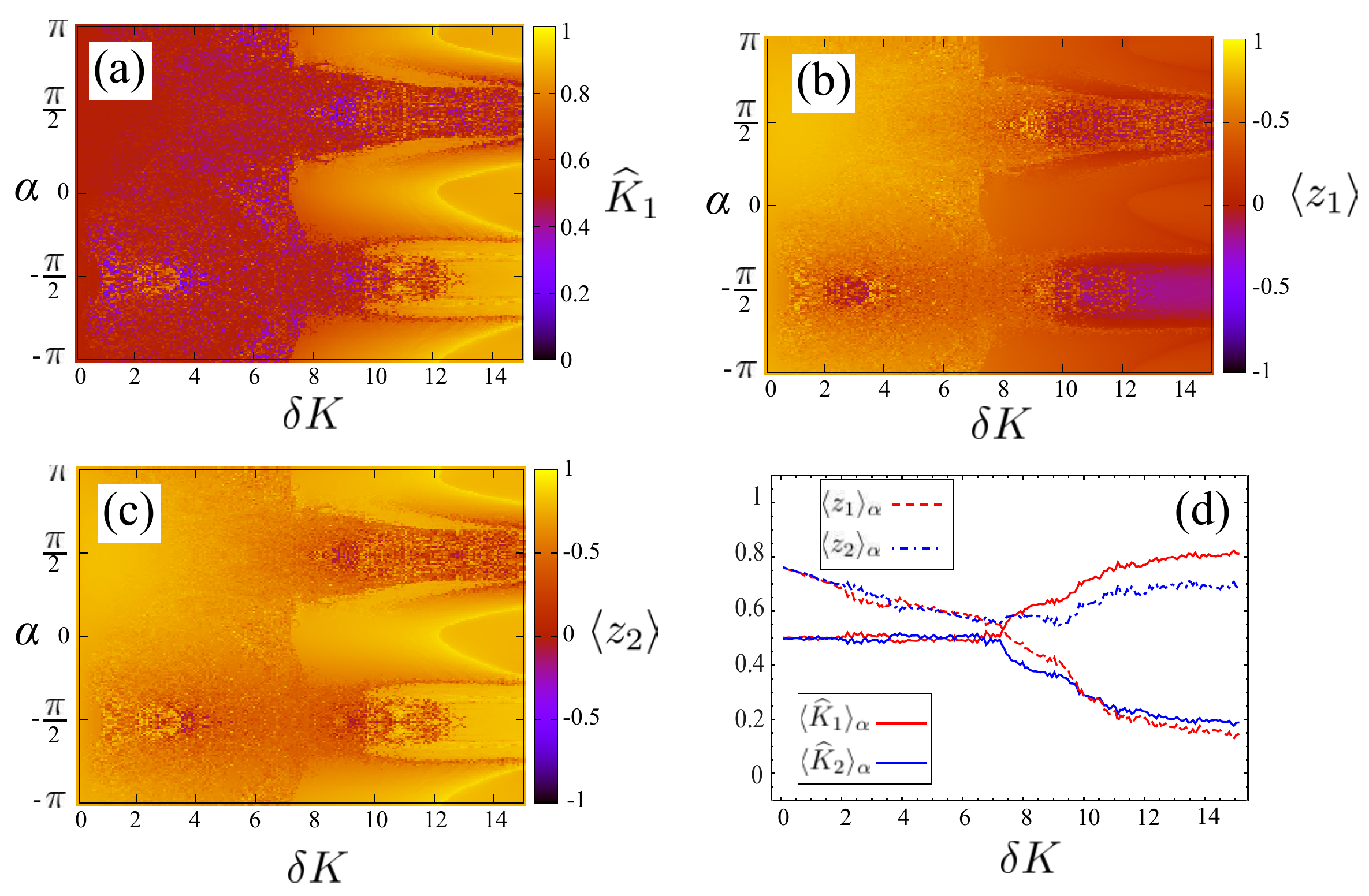}
\caption{(a) Normalized time-averaged kinetic energy $\widehat K_1$ of the first dipole. (b)-(c) Time
averages $\langle z_1 \rangle$ and $\langle z_2 \rangle$, respectively.
(d) $\langle \widehat K_i \rangle_{\alpha}$ and $\langle z_i \rangle_{\alpha}$
averaged with respect to the internal distribution angle $\alpha$.
$\beta=10$ holds for all the figures.}
\label{fi:Mapa4E10}
\end{figure}

Now, we increase the electric field strength  up to $\beta=10$, \ie,  the electric field interaction 
is one order of magnitude larger than the dipole-dipole one, the corresponding results are
presented in~\autoref{fi:Mapa4E10}. In the color map of~\autoref{fi:Mapa4E10}(a), we encounter
that most of the initial conditions lead the systems to equipartition for $\delta K\lesssim 7$.
Analogously to the $\beta=1$ dynamics, a more complex behavior appears for $\delta K\gtrsim 7$. 
Except for a region around $\alpha=\pi/2$, the energy transfer mechanism  is dominated by a 
non-equipartition regime, and most of the initial excess energy $\delta K$ remains in dipole one.
Although in this region around $\alpha=\pi/2$ the equipartition regime is dominant, there exist
regions of initial conditions that lead to non-equipartitioning.

The time-averaged inclinations $\langle z_1\rangle$ and $\langle z_2\rangle$ shown in 
Figs.~\ref{fi:Mapa4E10}(b)-(c) present similar patterns as those observed for $\widehat K_1$ 
in~\autoref{fi:Mapa4E10}(a). In the region dominated by energy equipartition 
$\delta K\lesssim 7$, the maps of  Figs.~\ref{fi:Mapa4E10}(b)-(c) indicate that the dipoles 
are significantly oriented along the electric field axis. For $\delta K\gtrsim 7$, the orientation 
of dipole one in~\autoref{fi:Mapa4E10}(b) decreases as $\delta K$ increases, and 
around $\alpha=-\pi/2$, there are initial conditions leading even to antiorientation,
which correspond to this dipole having most of the energy in~\autoref{fi:Mapa4E10}~(a). 
In contrast, the orientation of dipole two is globally enhanced 
in the non-equipartition region $\delta K\gtrsim 7$, see~\autoref{fi:Mapa4E10}(c). Only for initial 
conditions in the region around $\alpha = \pi/2$  both dipoles have similar orientations slightly 
directed towards the electric field. In other words, when the dipole one begins to rotate in the 
direction of the electric field and equipartition is dominant, the dipoles end up having enough 
energy to prevent their orientation. That is, for initial conditions around $\alpha = \pi/2$, the 
non-equipartition regime existing in the absence of the electric field $\beta=0$ for 
$\delta K\gtrsim 6$, see~\autoref{fi:Mapa1E0},  is mostly broken when the field is switched on. In 
contrast, for initial conditions out of that region, the system tends to remain in the same 
non-equipartitioning regime exhibited in the absence of the electric field, with most of the energy in 
dipole one. Thus, the electric field interaction is still not  able to relax the system out of the 
non-equipartition regime, and, therefore, only the second dipole is significantly oriented by
the electric field.

Again, the $\delta K$-evolution of the $\alpha$-averaged quantities in~\autoref{fi:Mapa4E10}(d)
confirms the behavior of the time-averaged ones presented in Figs.~\ref{fi:Mapa4E10}(a)-(c).
During the equipartition regime for $\delta K\lesssim 7$  the  dipoles have similar orientation. 
This orientation is maximal for small values of $\delta K$ and decreases as  $\delta K$ increases.
For $\delta K\gtrsim 7$, $\langle z_1 \rangle_{\alpha}$ continues decreasing because most of the energy 
is stored in dipole one, whereas  $\langle z_2 \rangle_{\alpha}$ grows slightly for increasing $\delta K$.  

\begin{figure}[t]
\includegraphics[width=.99\linewidth]{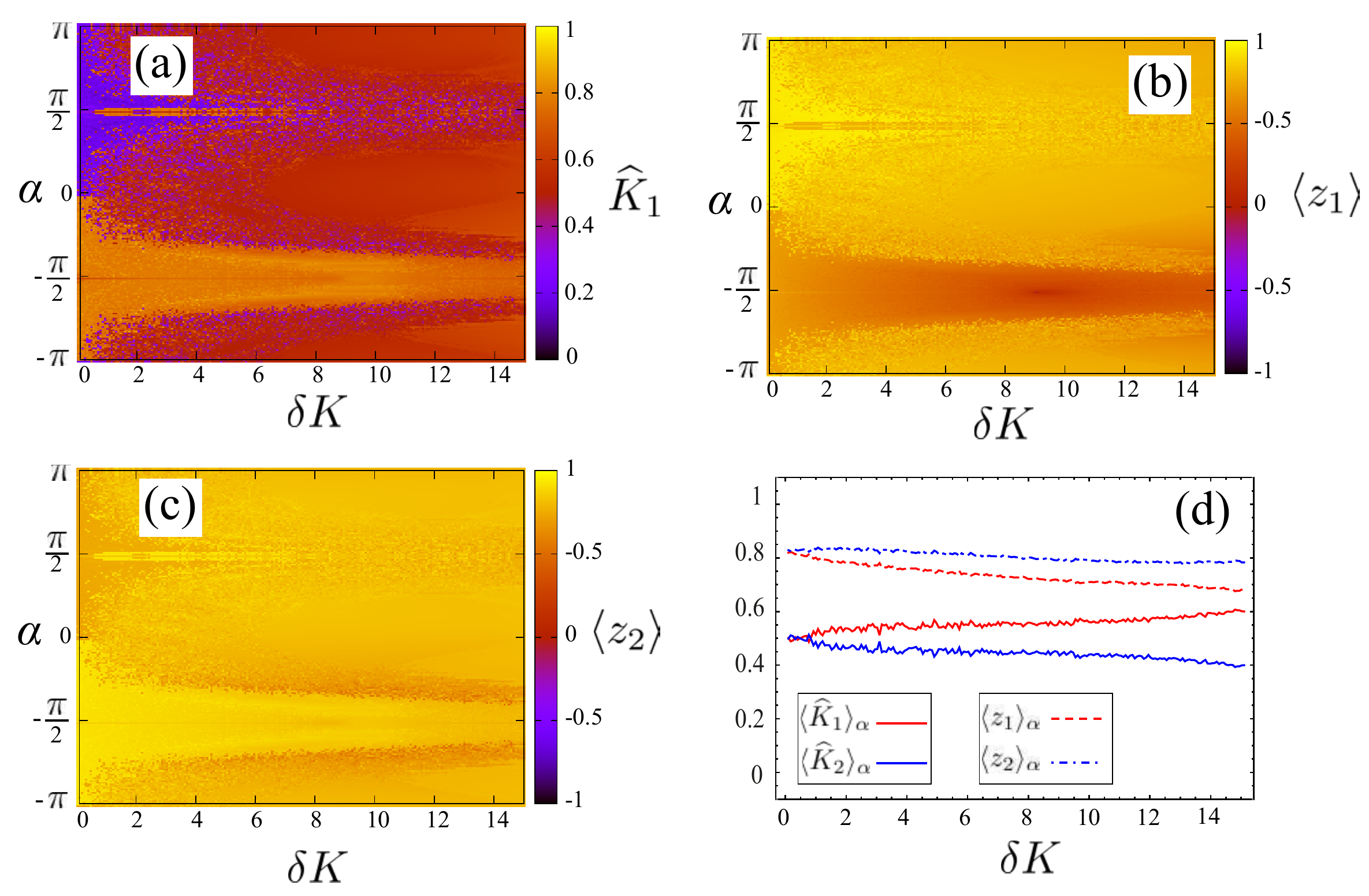}
\caption{(a) Normalized time-averaged kinetic energy $\widehat K_1$ of the first dipole. (b)-(c) Time
averaged $\langle z_1 \rangle$ and $\langle z_2 \rangle$, respectively.
(d) $\langle \widehat K_i \rangle_{\alpha}$ and $\langle z_i \rangle_{\alpha}$
averaged with respect to the internal distribution angle $\alpha$. 
All the panels are for an electric field parameter $\beta=100$.}
\label{fi:Mapa5E100}
\end{figure}

By further increasing the electric field to $\beta=100$, the dynamics is dominated by the interaction 
with the electric field. In~\autoref{fi:Mapa5E100}(a), the normalized time-averaged kinetic energy 
$\widehat K_1$ shows a complex behavior strongly depending on the initial excess energy 
$\delta K$ and on the angle $\alpha$. The map of~\autoref{fi:Mapa5E100}(a) shows wide 
regions of energy equipartition. Interestingly, there is a blue-colored (brown-colored) area centered
around $\alpha=\pi/2$  ($\alpha=-\pi/2$) where most of the kinetic energy is stored in dipole two (one). 
This means that when dipole one starts to rotate on an axis perpendicular to its $x_1z_1$ plane, the 
equipartition regime is not reached. 
For this strong electric field, a substantial orientation of the dipoles is expected, which is confirmed 
by the large values of $\langle z_1\rangle$ and $\langle z_2 \rangle$ in~\autoref{fi:Mapa5E100}(b)-(c).
Regardless of this orientation, Figs.~\ref{fi:Mapa5E100}(b) and (c) present similar structures 
as~\autoref{fi:Mapa5E100}(a), as was previously observed for $\beta=1$ and $10$. 
Thus, the non-equipartition regions in~\autoref{fi:Mapa5E100}(a) with most of the kinetic 
energy located in one of the dipoles correspond to the lighter and darker red colored regions 
of Figs.~\ref{fi:Mapa5E100}(b)-(c), where the orientation is minimal and maximal, respectively.
The  $\alpha$-averaged $\langle \widehat K_1 \rangle_{\alpha}$
and $\langle \widehat K_2 \rangle_{\alpha}$ of~\autoref{fi:Mapa5E100}(d) confirm that
the system is always near equipartition, with a small positive energy balance for dipole one.
Due to the strong electric field $\beta=100$, $\langle \widehat K_1 \rangle_{\alpha}$ 
and $\langle \widehat K_1 \rangle_{\alpha}$ show just a slight variation as  $\delta K$ increases.
A similar situation is found for the $\alpha$-averaged $\langle z_i \rangle_{\alpha}$, where
the large orientation induced by the field is only slightly counteracted for increasing initial energy $\delta K$.
Again, the less energetic dipole  two is more oriented than the more  energetic dipole one.

\begin{figure}[t]
\includegraphics[width=.99\linewidth]{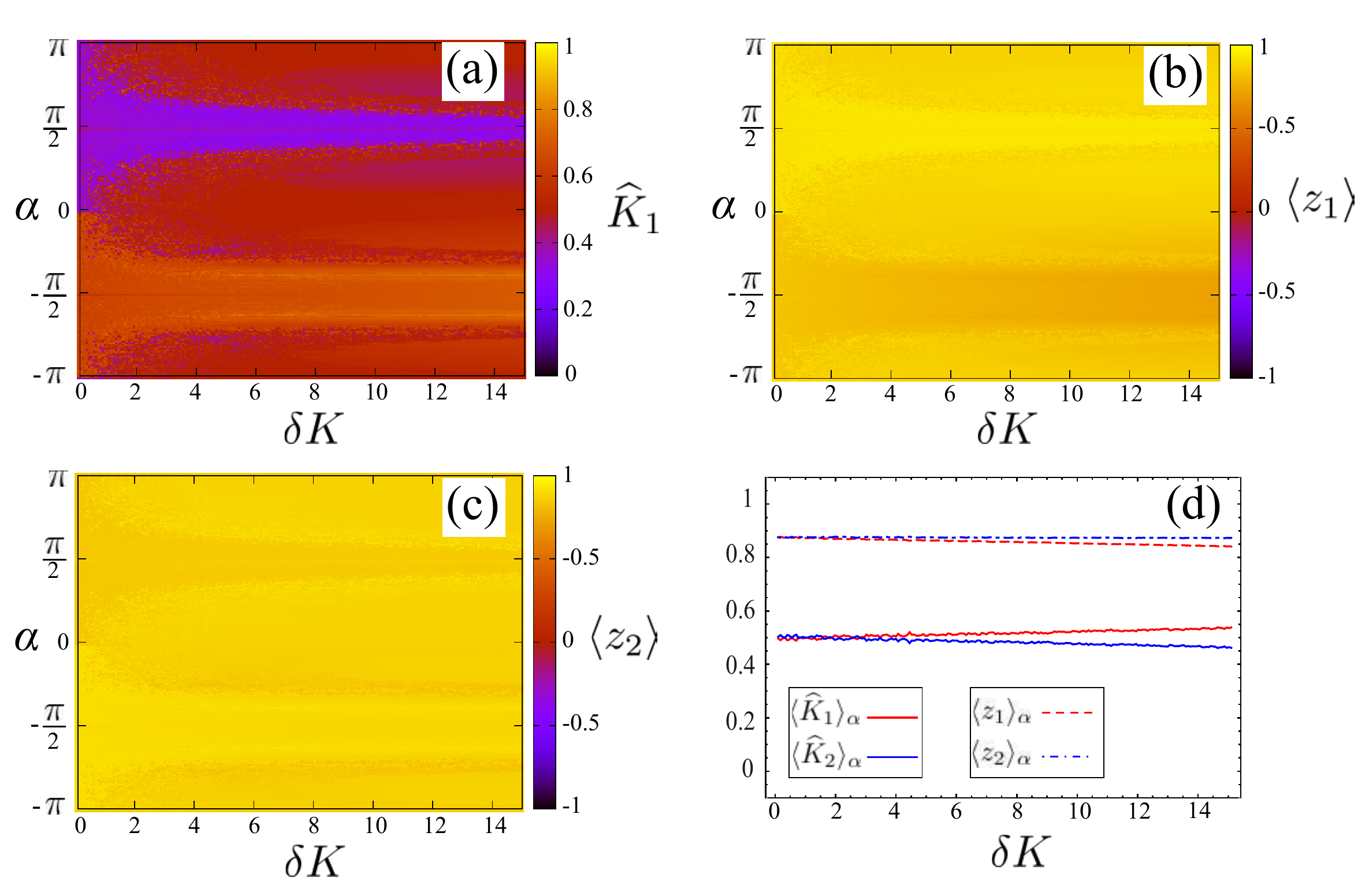}
\caption{(a) Normalized time-averaged kinetic energy $\widehat K_1$ of the first dipole. (b)-(c) Time
averages $\langle z_1 \rangle$ and $\langle z_2 \rangle$, respectively.
(d) $\langle \widehat K_i \rangle_{\alpha}$ and $\langle z_i \rangle_{\alpha}$
averaged with respect to the internal distribution angle $\alpha$.
All the panels are for an  electric field parameter $\beta=1000$.}
\label{fi:Mapa6E1000}
\end{figure}
When the electric field  interaction is three order of magnitude larger than the dipole-dipole one \ie, 
$\beta=1000$, most of the initial conditions leave the
system close to the equipartition regime see~\autoref{fi:Mapa6E1000}(a). 
Moreover, the normalized time-averaged kinetic energy $\widehat K_1$  and the orientations 
show similar behaviours as those for the $\beta=100$ case, compare 
Figs.~\ref{fi:Mapa5E100}(a)-(b)-(c) and Figs.~\ref{fi:Mapa6E1000}(a)-(b)-(c).
We again encounter the non-equipartition regions around $\alpha=\pi/2$ and $-\pi/2$, where most of the kinetic 
energy is stored  in dipole two and one, respectively. 
Thus, the energy equipartition regime is closely related to a smaller orientation of dipole two and viceversa. 
However, due to the strong electric field the two dipoles are significantly oriented along the electric 
field direction, see Figs.~\ref{fi:Mapa6E1000}(b)-(c). The $\alpha$-averaged 
$\langle \widehat K_i\rangle_{\alpha}$ and $\langle z_i \rangle_{\alpha}$ of~\autoref{fi:Mapa6E1000} 
assert that the system is always close to the equipartition regime, whereas the strong field 
 orients the dipoles with $\langle z_i \rangle_{\alpha}>0.8$ with $i=1,2$.

\section{Regular and chaotic dynamics}
\label{sec:chaos}
In this section,  we  study the chaoticity of the  dynamics and its possible relation to the 
time-averaged energy transfer. The chaotic character of an orbit in a dynamical system is 
related to its sensitivity to the corresponding initial conditions. As a numerical measure of this sensitivity, we use 
the so-called, Fast Lyapunov Indicator (FLI)~\cite{Froeschle1,Froeschle2,Fouchard} to determine 
the degree of chaoticity of an orbit. Given a $n-$dimensional flow of a dynamical system 
\begin{equation}
\label{ecumovi}
\frac{d {\bf r}}{dt}={\bf f}({\bf r},t),
\end{equation}
the time evolution of the variational vector $\delta{\bf r}(t)$ is provided by the variational equation 
\begin{equation}
\label{ecuvari}
\frac{d \delta{\bf r}}{dt}=\frac{\partial{\bf f}({\bf r},t)}{\partial {\bf r}} \ \delta {\bf r}.
\end{equation}
For a given orbit with initial conditions ${\bf r(0)}$ and $\delta{\bf r(0)}$, the numerical integration of equations~\ref{ecumovi} 
and~\ref{ecuvari} up to a final time $t_f$ yields the value of the FLI of that orbit  defined by 
\begin{equation}
\label{ofli}
\mbox{FLI}({\bf r}(0), \delta{\bf r}(0), t_f)=\sup_{0\le t \le t_f}  \log \| \delta{\bf r}(t) \|.
\end{equation}
The variational vector $\delta{\bf r}$ increases linearly with time for regular periodic and quasiperiodic orbits, whereas it increases exponentially for chaotic orbits.

\begin{figure*}
\includegraphics[width=0.32\linewidth]{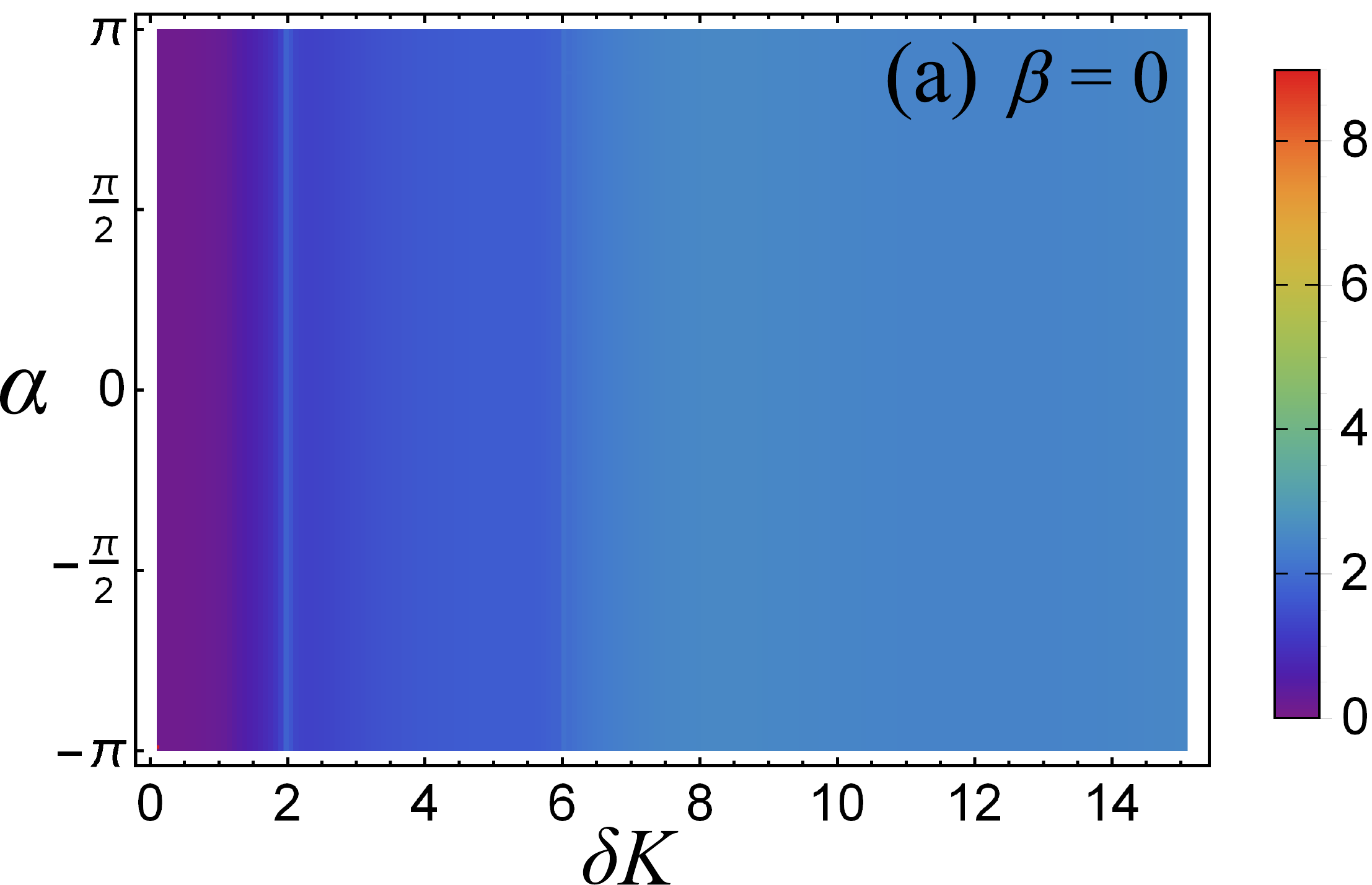}\,\includegraphics[width=0.32\linewidth]{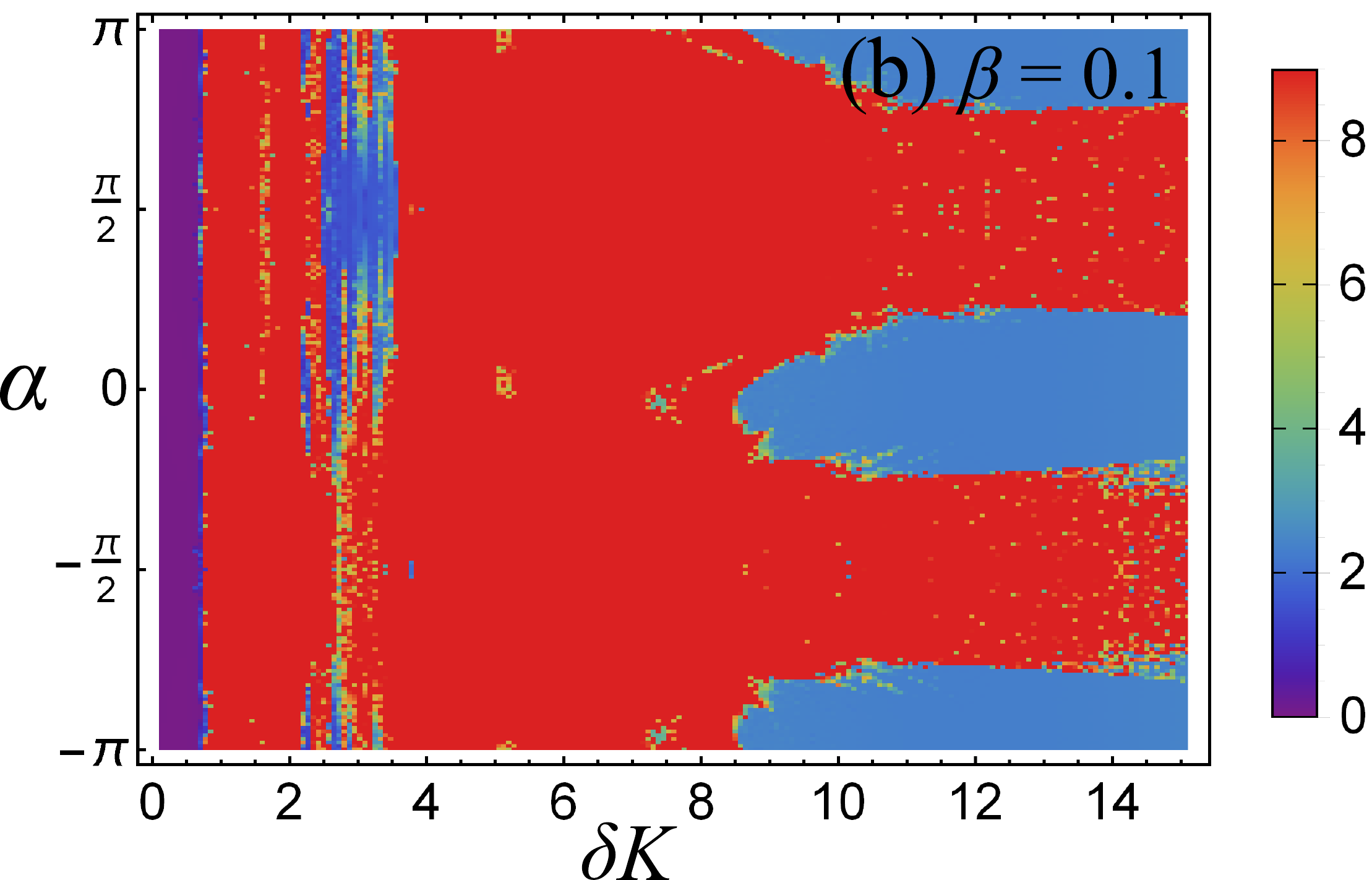}\,
\includegraphics[width=0.32\linewidth]{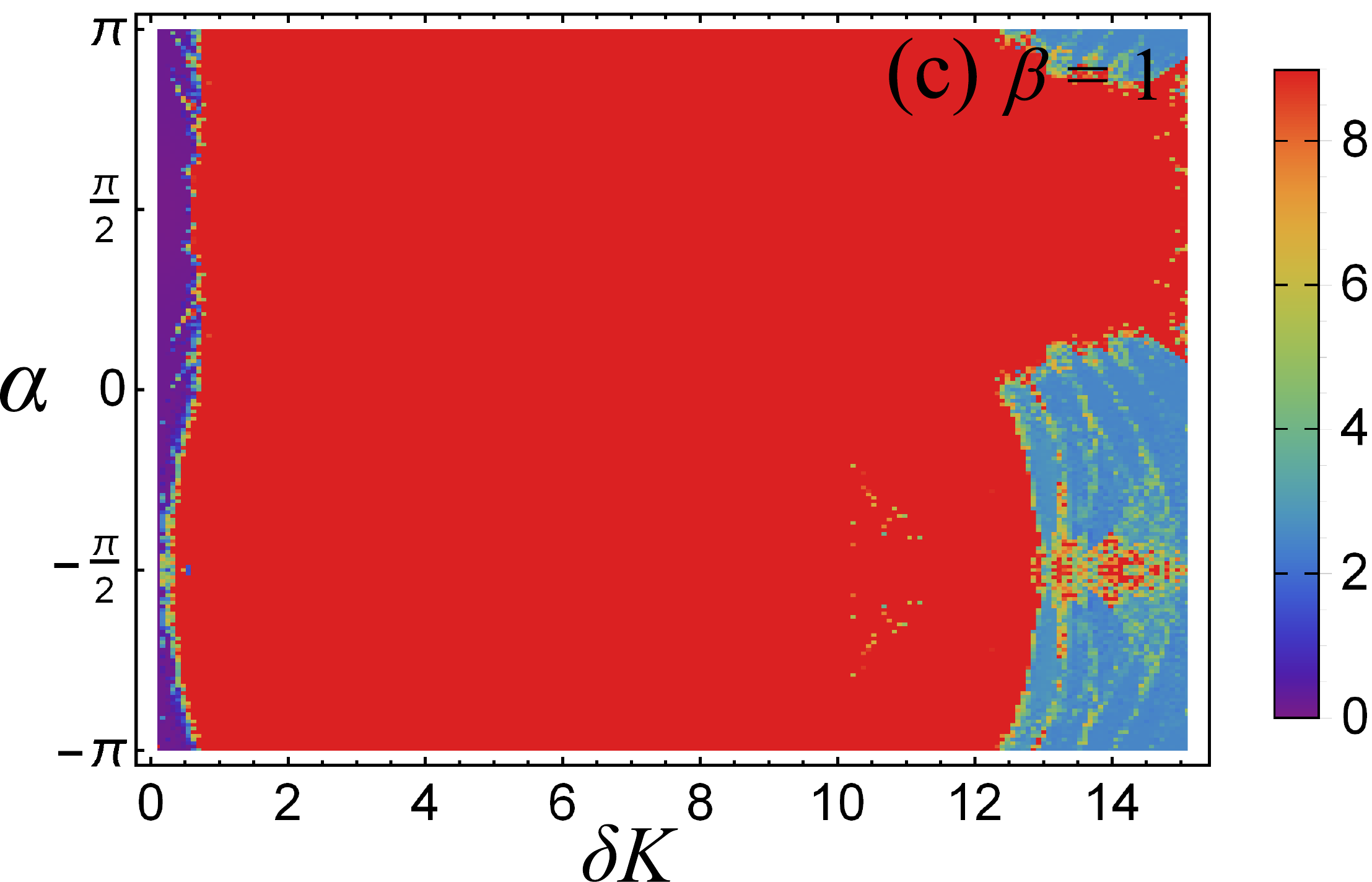}\\ \includegraphics[width=0.32\linewidth]{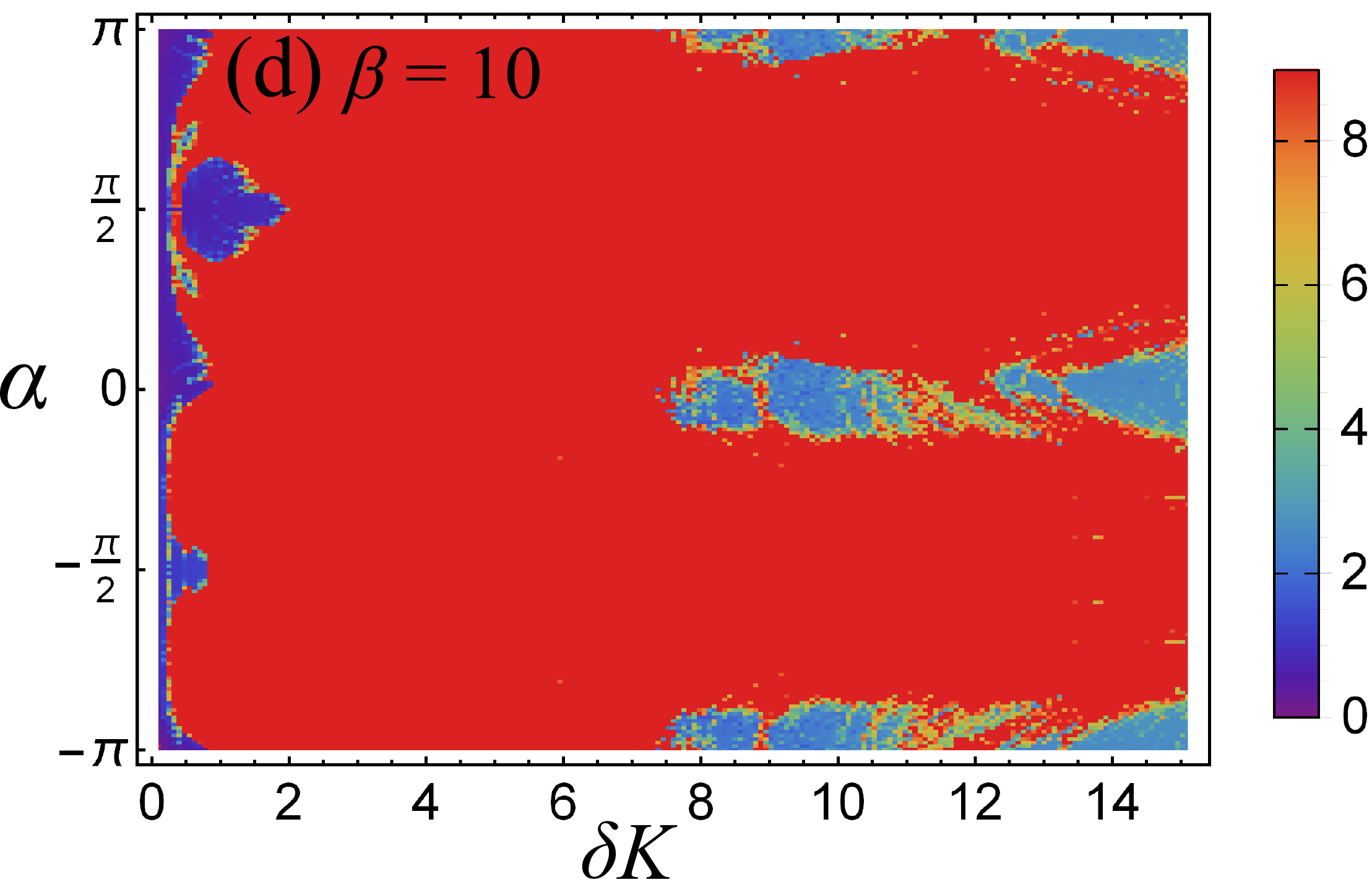}\,
\includegraphics[width=0.32\linewidth]{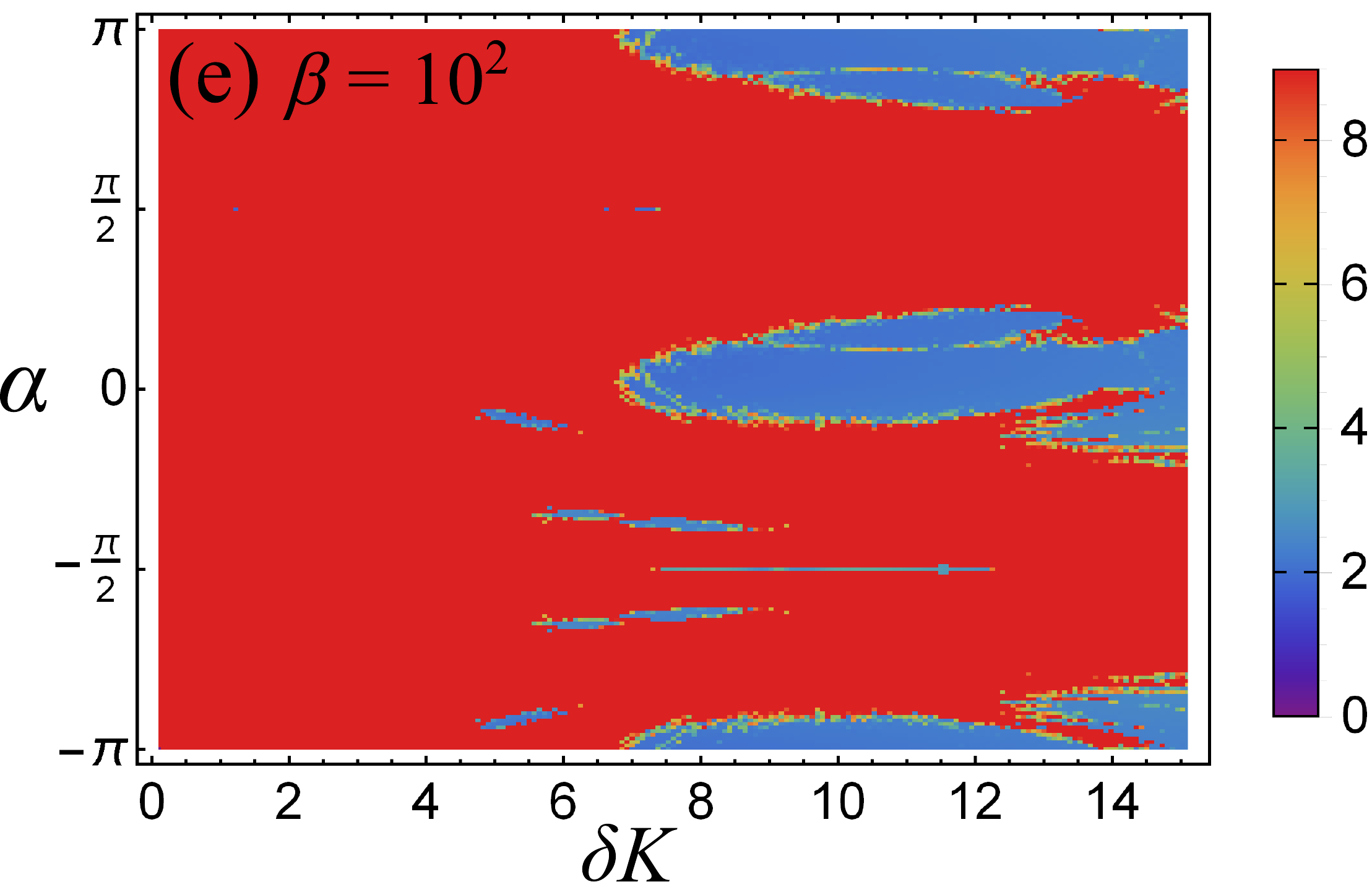}\, \includegraphics[width=0.32\linewidth]{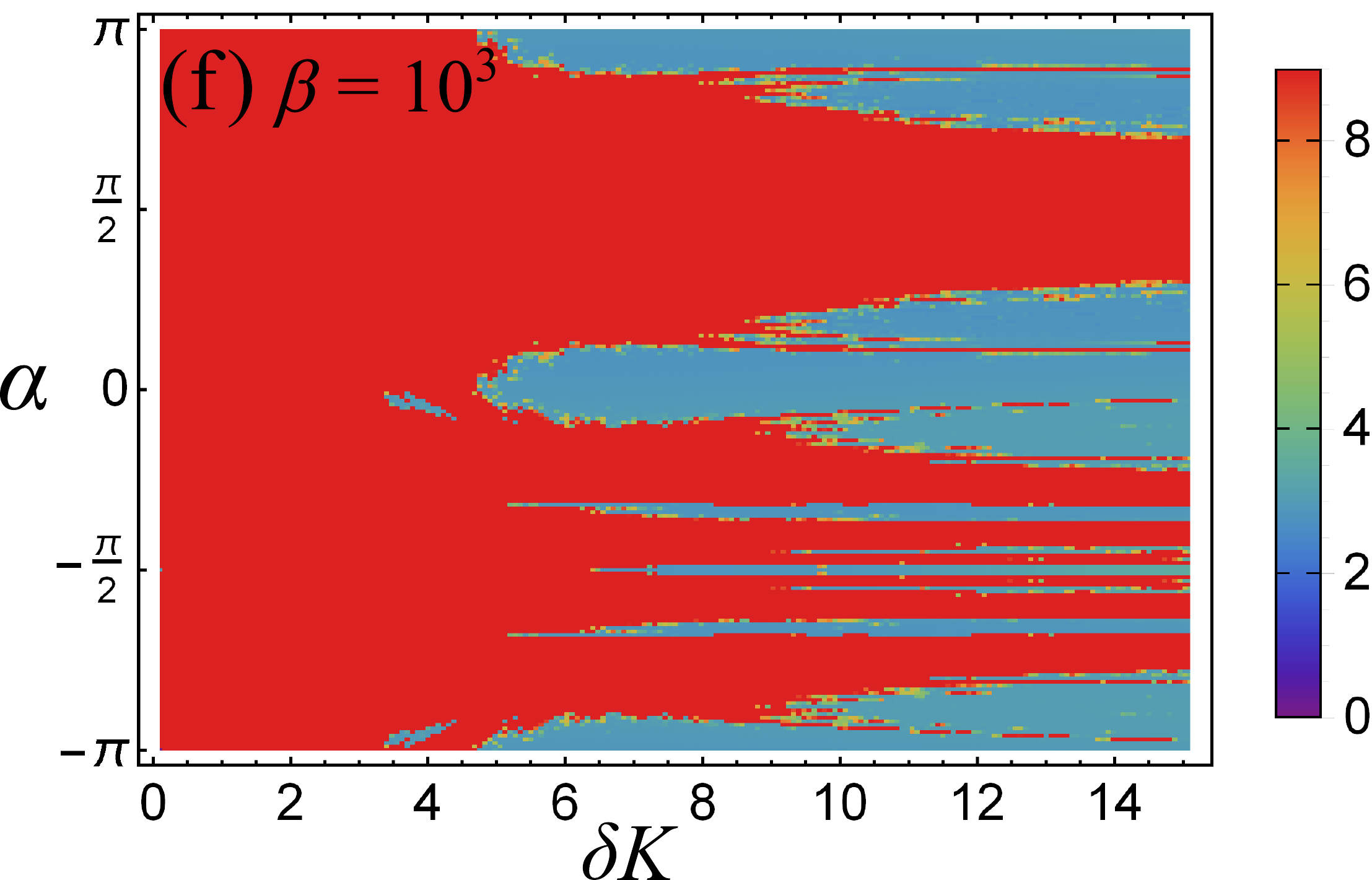}
\caption{FLI values for different initial conditions with $\delta {\cal K} \in [0.1, 15]$ and $\alpha \in [-\pi,\pi)$.}
\label{fi:FLImaps}
\end{figure*}

We have calculated FLI for varying  initial conditions $\delta K$ and  $\alpha$. 
To do so, we have integrated  the equations of motion and their first variational equations 
in Cartesian coordinates by means of the St\"ormer-Verlet algorithm~\cite{Andersen} using 
the same initial conditions as in the previous section. Now, we stop the integration if the FLI 
reaches the saturation value nine, that characterizes an orbit as chaotic, or if the integration 
time reaches the stopping value $t_f=3000$. Our numerical tests have shown that this 
stopping time and saturation limit are adequate to correctly characterize  any orbit.

The  FLI color maps are presented in~\autoref{fi:FLImaps}  for the electric field strengths
$\beta=0,0.1, 1, 10, 100$ and $1000$, the initial kinetic energy excess $\delta {\cal K}\in [0,15]$ 
and the angle $\alpha \in [-\pi,\pi)$. It is worth noting that, along the horizontal lines $\alpha=0$ 
and $\alpha=\pm \pi$ in~\autoref{fi:FLImaps},  the system evolves on the invariant manifold 
${\cal M}$ given by the planes $y_1=y_2=0$, \ie, both dipoles are restricted to rotate in the common  
$x_1 z_1\equiv x_2 z_2$ plane. 

\begin{figure*}
\includegraphics[width=0.4\linewidth]{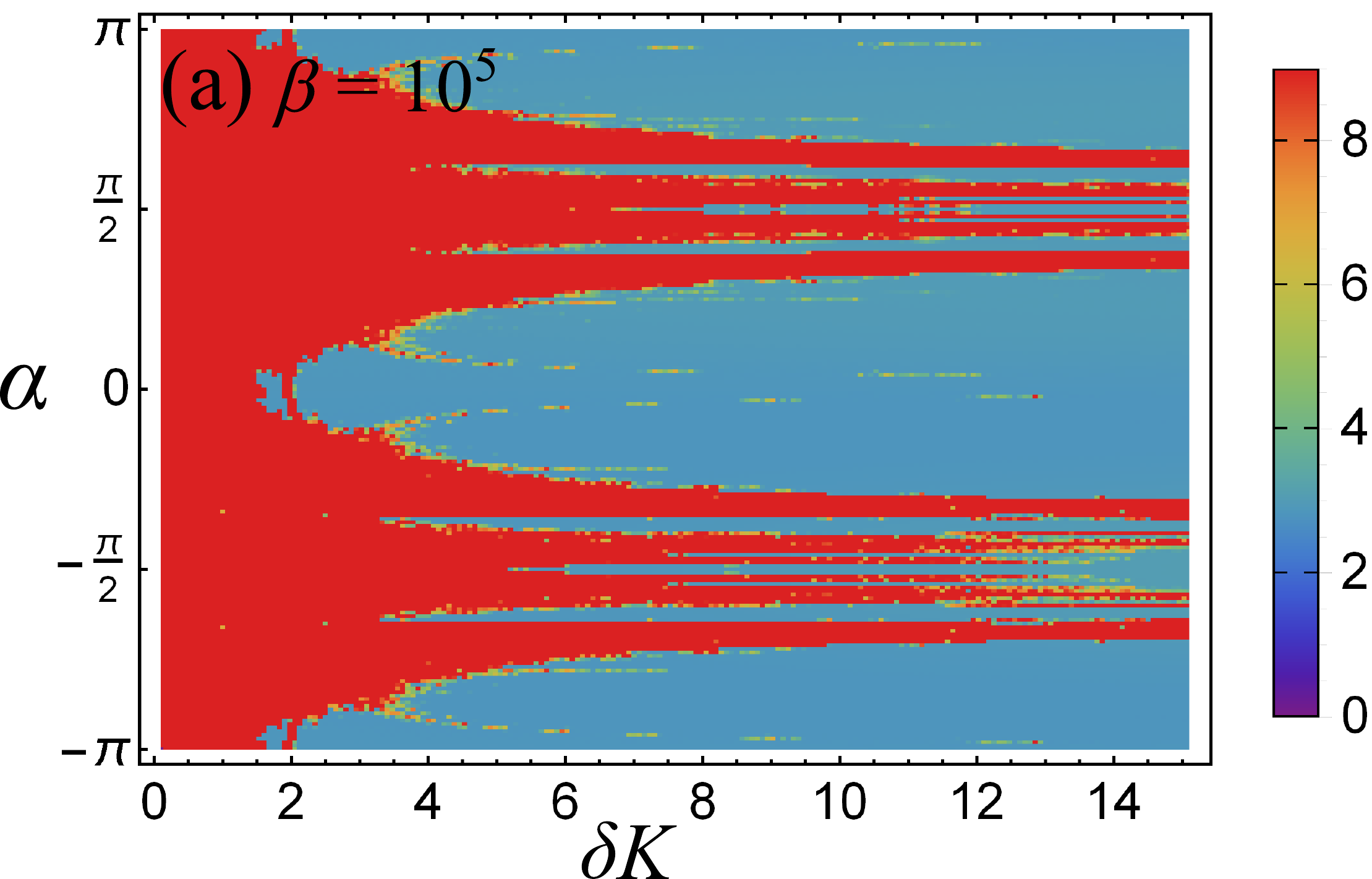}\,\includegraphics[width=0.4\linewidth]{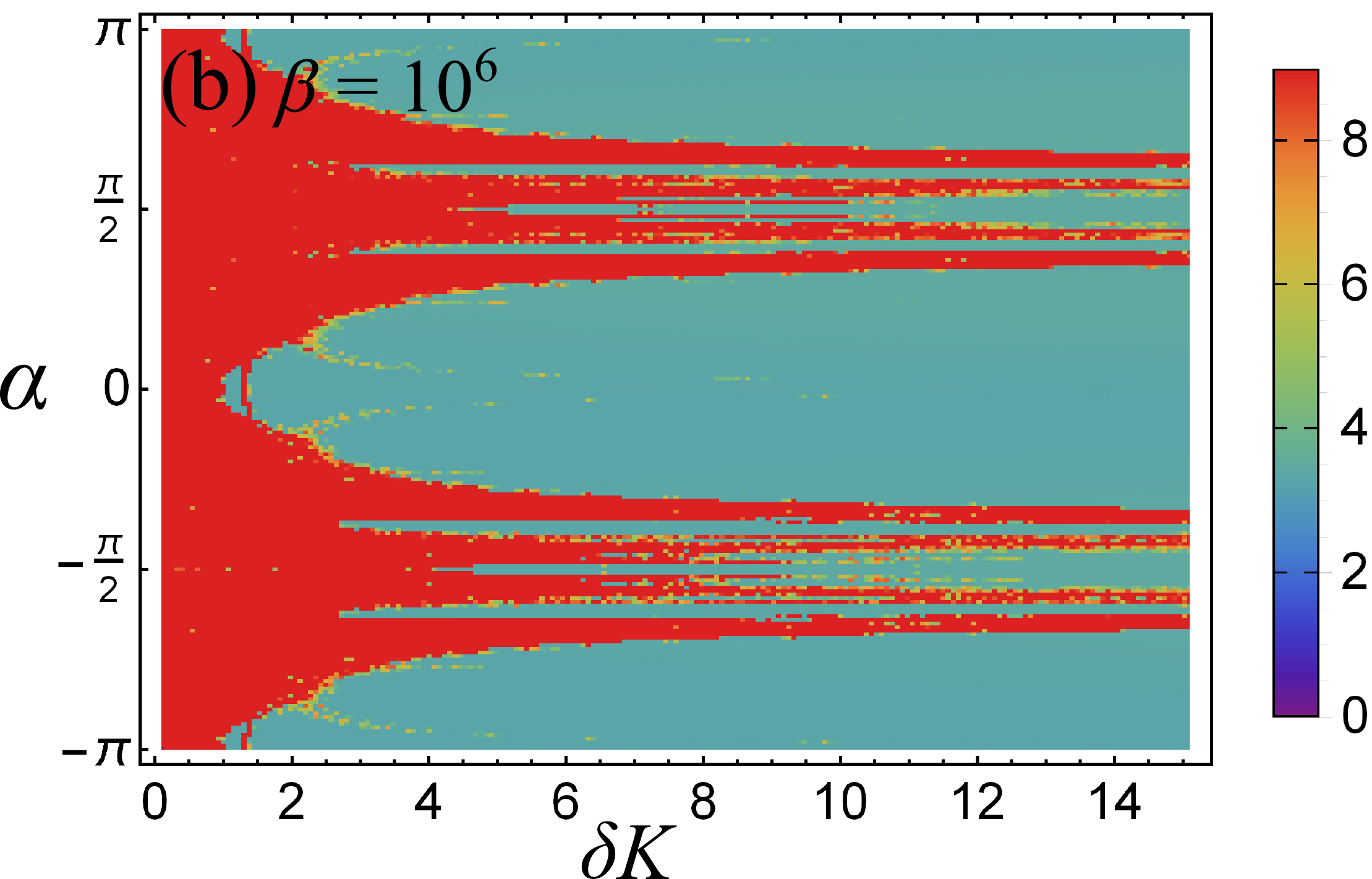}\,
\caption{FLI values for different initial conditions
with $\delta {\cal K} \in [0.1, 15]$ and $\alpha \in [-\pi,\pi)$ for (a) $\beta=10^5$ and
(b) $\beta=10^6$.}
\label{fi:FLImaps2}
\end{figure*}

In the absence of the electric field, $\beta=0$, we observe in~\autoref{fi:FLImaps}(a) that the
FLI values do not depend on the angle $\alpha$, that is, on the direction in which the first 
dipole begins to rotate. As explained above, this is because regardless of the value of $\alpha$, 
the system will always move in one of the equivalent invariant manifolds ${\cal M}$. Moreover, 
for zero electric field, the FLI value increases with the kinetic energy excess $\delta {\cal K}$, 
although the dynamics is regular for all initial conditions. By switching on  the electric field, 
most of the dynamics becomes chaotic even for weak electric fields such as $\beta=0.1$, 
see~\autoref{fi:FLImaps}(b). For electric field values up to $\beta=10$, the regular behavior only 
persists for low values of the kinetic energy excess $\delta {\cal K}<1$ and particularly around $\alpha=0$ 
and $\alpha=\pm \pi$, see Figs.~\ref{fi:FLImaps}(b-d).
For  stronger electric fields, the chaotic behavior of the system still persists in
wide regions of the FLI maps~\autoref{fi:FLImaps}(e-f) for  $\beta=$100 and 1000
and for the range of $\delta K$ values shown. This is
somehow an unexpected result because for
increasing field strength the dynamics is gradually dominated by the
interaction of the dipoles with the field. This would lead the system to a (quasi) integrable regime, where most of the
dynamics would be regular.
However, the dipolar interaction, although being small compared to the electric field interaction, it
is still able to 
cause a substantial volume of chaotic motion.
In this sense, we have calculated the FLI maps in the range $0<\delta K\le15$
for very large values of the field parameter, namely for
$\beta=10^5$ and $10^6$ (see Fig.\ref{fi:FLImaps2}). Indeed, we observe in the color
maps of Fig.\ref{fi:FLImaps2}. that, for increasing values of
$\beta$, the regions of initial conditions leading to regular motion grow in size while the regions of
initial conditions leading to chaotic orbits shrink. However, this global tendency to regular motion is very
slow, which confirms the relevant role that the DDI is playing in the dynamics even for very large electric field values.

A comparison of the results obtained for the time-averaged energy transfer, see 
Figs.~\ref{fi:Mapa2E01}-\ref{fi:Mapa6E1000},  with those for the degree of chaoticity presented 
in~\autoref{fi:FLImaps} does not show any significant correlation. This may be due to the fact 
that regular and chaotic orbits can achieve the same time-averaged kinetic energy.  We have 
additionally computed  another chaos indicator, the so called Smaller Alignment Index 
(SALI)~\cite{Skokos1,Skokos2}, and the corresponding results are in perfect agreement with the 
FLI ones presented here.

\section{Conclusions}
\label{sec:con}

We have theoretically investigated two interacting classical  dipoles fixed in 
space, which are described as rigid rotors, in the presence of a homogenous 
electric field. The two dipoles are initially at rest in the stable head-tail configuration.
Initially, one of the dipoles is pushed out of this stable configuration by giving it a specific velocity 
in a direction perpendicular to the head-tail axis, and at the same time the electric field is switched on. 
The following classical dynamics is explored in terms of the energy transfer 
mechanisms between the two dipoles and  their orientations along the electric field axis.

The field-free dynamics of this system was previously investigated in the invariant manifold 
${\cal M}_{XZ}$~\ref{manifoldMxz} in Ref.~\cite{rotors2D}. Here, we have shown that 
the dynamics is independent of the initial rotation angle of the dipole because it is
 restricted to one of the invariant manifolds of the infinite set ${\cal M}$. Depending on the 
initial excess energy given to one of the dipoles, the system falls to either an energy 
equipartition regime or a non-equipartition one, and the dynamics is regular.

The classical field-dressed dynamics strongly depends on the electric field strength and on the 
initial conditions. For weak external fields, the dynamics is still dominated by the dipole-dipole 
interaction and the energy transfer dynamics resembles the field-free dynamics with the two 
dipoles having a small orientation. By increasing the electric field, the interaction with this field 
dominates the classical dynamics. The size of the non-equipartition region appearing
for large values of the initial excess energy decreases as the electric field increases,
and the system tends to the energy equipartition regime. In addition, the orientation of the two 
dipoles along the electric field direction increases. We note that when the initial momentum is  almost 
parallel of antiparallel to the electric field direction, we observe that even for very strong electric 
fields, the  non-equipartition behavior dominates the dynamics.
Finally, we find that, even for large electric 
field strengths and for our considered excess energies, the system shows a highly chaotic behavior.
This is a remarkable feature because for
increasing electric field strength it would be expected that the system gradually tends to a (quasi) integrable
state where the dynamics is fully dominated by the
 interaction of the dipoles with the field and where the mutual dipole interaction would be considered
 as a small perturbation.
However, as we have observed for very large values of the electric field, the DDI is
still able to generate significant regions of chaotic motion.

A natural continuation of this work would be the investigation of the energy transfer and the possible
collective phenomena
of systems with many 3D-dipoles including in particular, the study of a linear chain of 3D-dipoles.

\begin{acknowledgments}  

M.I. and J.P.S. acknowledge financial
support by the Spanish Project No. MTM 2017-88137-C2-2-P (MINECO).
R.G.F. gratefully acknowledges  financial support by the Spanish Project No. FIS2017-89349-P 
(MINECO), and by the Andalusian research group FQM-207. This study has been partially 
financed by the Consejer\'{\i}a de Conocimiento, Investigaci\'on y Universidad, Junta de 
Andaluc\'{\i}a and European Regional Development Fund (ERDF), Ref. SOMM17/6105/UGR. 
\end{acknowledgments}

\appendix
\section{Stability, existence and bifurcations of the equilibrium points}
Here, we discuss  the existence, stability and energy of the equilibria of the systems formed by two 
dipoles in an external electric field. These equilibria are  summarized in~\autoref{ta:tabla1}. 
We also analyze the bifurcations appearing between these equilibria as the electric field strength is 
increased.

\begin{table*}
\caption{Conditions of existence, stability and energy of the critical points of 
${\cal V}_1( \mathbf{l}_1, \mathbf{l}_2)$ in~\autoref{potencarte}. The acronyms SPR1, SPR2 and 
SPR3 denote saddle points of rank-one, rank-two and rank-three, respectively.}
\centering
\begin{adjustbox}{max width=\textwidth}
\begin{tabular}{*{4}{|l|c|l|l}|}
%\hline\noalign{\smallskip}
\hline
Equilibria & Existence & Stability & Energy ${\cal E}$  \\[2ex]
\hline
$D_1$  & $\beta=0$ & Degenerate circle of equilibria & $E_{D_1}=1$ \\[2ex]
$D_2$  & $\beta=0$ & Degenerate circle of equilibria & $E_{D_2}=-1$ \\[2ex]
$P_1$ & $\beta < 3$ & Minima &
$E_1 = -(6+\beta^2)/3$ \\[2ex]
$P_2$ & $\beta>0$ & SPR2 & $E_2 = -1$ \\[2ex]
$P_3$ & $\beta>0$ & If $0<\beta < 2$: SPR2; if $2 < \beta < 3$: SPR1; If $\beta > 3$: Minima
 &$E_3 =  1 - 2 \beta$\\[2ex]
$P_4$ & $\beta>0$ & If $0<\beta < 1$: SPR3; if $\beta > 1$: Maxima & $E_4 =  1 + 2 \beta$\\[2ex]
$P_5$ & $\beta < 1$ & Maxima &
$E_5 = (2 +\beta^2)$ \\[2ex]
$P_6$ & $0<\beta< 2$ & SPR1 &
$E_6=-(2+\beta^2)/2$ \\[2ex]
\hline
\end{tabular}
\end{adjustbox}
\label{ta:tabla1}
\end{table*}

\subsection{Stability and existence of the equilibria for $\beta = 0$}
For zero electric field, $\beta=0$, the only isolated equilibria are $P_1$ and $P_5$. 
Their nature is deduced by applying the method of Lagrange multipliers 
%of a function subject to several constraints.
%In our case, such a function is 
to the potential
energy surface ${\cal V}_1(\mathbf{l}_1, \mathbf{l}_2,t=0)$ ~\autoref{potenCarte3} with $\beta=0$, 
imposing  the constraints $\{\mathbf{l}_i=(x_i, y_i, z_i) ||\, |\mathbf{l}_i|^2=1 \}_{i=1}^2$.
There exist the two stable  equilibria $P_1$ with minimal energy $E_1=-2$, which are minima, and
the two   unstable equilibria $P_5$ with  maximal energy $E_5=2$,  which are maxima. 

\subsection{Stability and existence of the equilibria for $\beta \ne 0$ and $t\ge t_1$}
For the general case $\beta \ne 0$, instead
of using Cartesian variables, the analysis of the stability of the equilibria is carried out in spherical
variables $\{(\theta_i,P_{\theta_i}),(\phi_i,P_{\phi_i})\}_{i=1}^2$.
The equilibrium points of the Hamiltonian flux~\ref{flux2} for $t\ge t_1$ are the critical
points of the potential 
${\cal V}_2$  [see~\autoref{potenEuler}], 
\begin{eqnarray}
%\label{potenEuler}
&{\cal V}_2(\theta_1, \phi_1, \theta_2, \phi_2, t\ge t_1)=
-\beta  (\sin \theta_1 \sin \phi_1+\sin \theta_2 \sin \phi_2)\nonumber\\[2ex]
 &+ [\sin\theta_1\sin\theta_2 \cos(\phi_1-\phi_2)-2 \cos\theta_1\cos\theta_2],\nonumber
 \end{eqnarray}
\noindent
together with the conditions $P_{\theta_i}=P_{\phi_i}=0$.
Obviously, the potential ${\cal V}_2(\theta_1, \phi_1, \theta_2, \phi_2, t\ge t_1)$ presents the same
critical points as ${\cal V}_1( \mathbf{l}_1, \mathbf{l}_2)$ in~\autoref{potencarte}. Thence, after expressing
in spherical coordinates the (isolated) critical points
$\{P_i\}^6_{i=1}$ detailed in~\autoref{sec:equilibria}, their stability
can be inferred from the nature of the corresponding eigenvalues of the Hessian matrix associated to
${\cal V}_2 (\theta_1, \phi_1, \theta_2, \phi_2, t\ge t_1)$.

\subsubsection{The equilibria $P_1$}
For the two equilibria $P_1$, the eigenvalues of the Hessian matrix are:
\[
\left(\beta ^2/9 , \ \beta ^2/3 , \ (9 - \beta ^2)/3 , \ (\beta ^2+3)/3 \right).
\]
Because the equilibria $P_1$ only exist when $0<\beta<3$, its four eigenvalues are positive, which indicate that $P_1$ are minima. In other words, when they exist, equilibria $P_1$ are stable. 
 
\subsubsection{The equilibria $P_2$}
The eigenvalues of the Hessian matrix of the two equilibria $P_2$ are:
\[
\left(1 -\sqrt{\beta ^2+1}, 1+ \sqrt{\beta ^2+1},1 -\sqrt{\beta ^2+4}, 1+ \sqrt{\beta ^2+4}\right).
\]
These eigenvalues indicate that $P_2$ are always rank-two saddle points because they have two positive and
two negative eigenvalues. Then, equilibria $P_2$ are always unstable.

\subsubsection{The equilibrium $P_3$}
The eigenvalues of the Hessian matrix of the equilibrium $P_3$ are:
\[
\left(\beta ,\ \beta +1, \ \beta -3, \ \beta -2  \right).
\]
The first and second eigenvalues are always positive.
For $0<\beta <2$,  the equilibrium $P_3$ is a rank-two saddle point because it has two positive and
two negative eigenvalues. In the interval $2 < \beta < 3$,  $P_3$ is a unstable rank-one saddle
point (the third eigenvalue is positive). For $\beta >3$, all the eigenvalues 
are positive, e.g., the equilibrium $P_3$ is a (stable) minimum.

\subsubsection{The equilibrium $P_4$}
The eigenvalues of the Hessian matrix of the equilibrium $P_4$ are:
\[
\left(-\beta -2, \ -\beta , \ -\beta -3,\ 1 -\beta \right).
\]
The first, second and third eigenvalues are always negative, while the fourth one 
changes from positive to negative for $\beta >1$. Then, $P_4$ is an unstable rank-three saddle point
when $0<\beta <1$, and a (unstable) maximum for $\beta >1$.

\subsubsection{The equilibria $P_5$}
The eigenvalues of the Hessian matrix of the two equilibria $P_5$ are:
\[
\left(-3 \beta ^2, \ -\beta ^2, \ -\beta ^2-3, \ \beta ^2-1 \right).
\]
Because equilibria $P_5$ only exist for $\beta <1$,  all the eigenvalues are negative, such 
that $P_5$ are always (unstable) maxima.

\subsubsection{The equilibria $P_6$}
The eigenvalues of the Hessian matrix of the two equilibria $P_6$ are:
\[
\left(\beta^2/2, \ 3, \ -1, \ (-\beta^2 + 4)/2 \right).
\]
The first and second eigenvalues are positive, the third one is  negative and the
fourth one is positive for $0<\beta < 2$.
Because equilibria $P_6$ only exist for $0<\beta <2$,  they are unstable rank-one saddle points.

\begin{figure}
\includegraphics[width=0.99\linewidth]{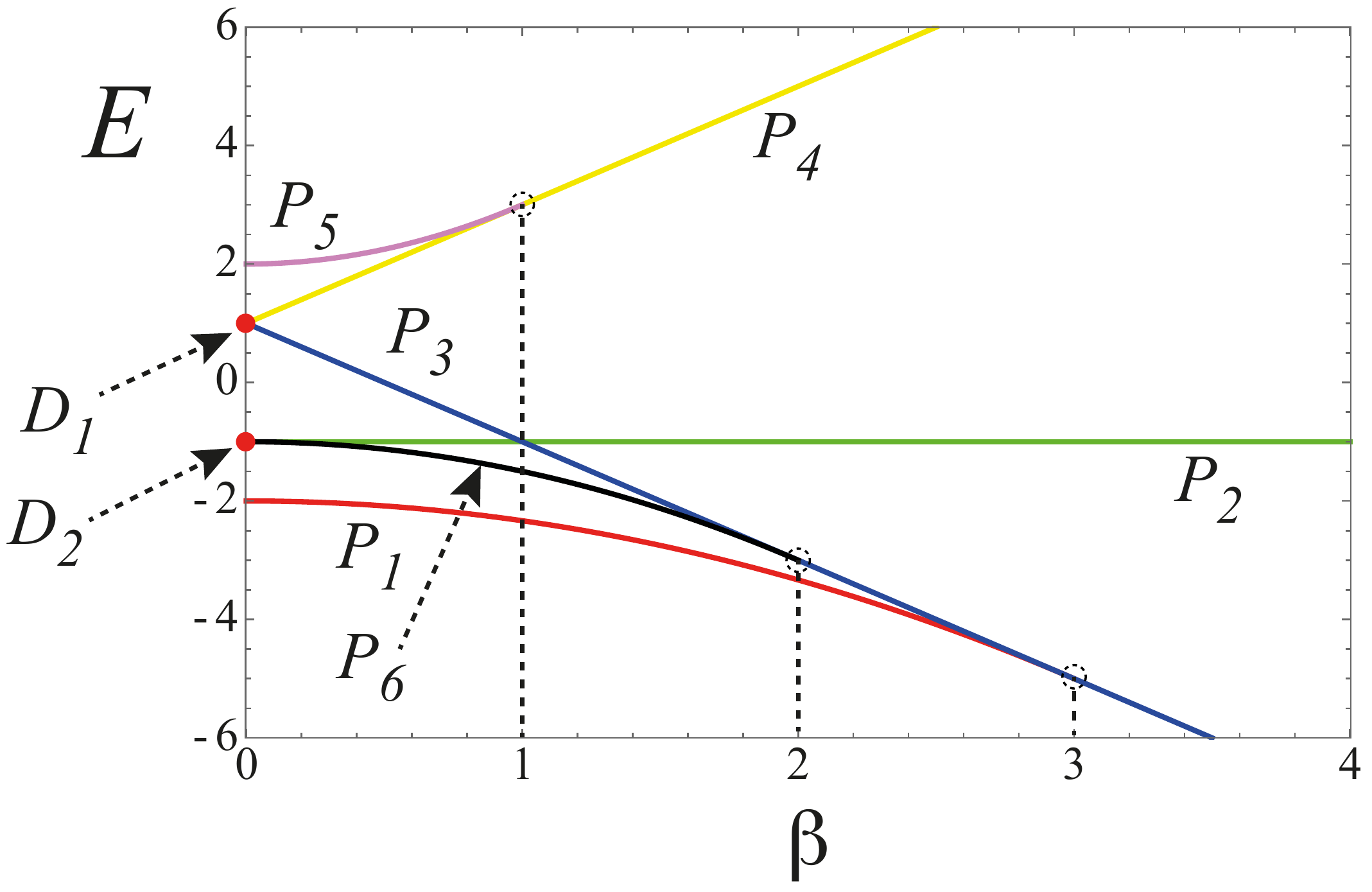}
\caption{Evolution of the energies of the critical points
of ${\cal V}_1( \mathbf{l}_1, \mathbf{l}_2)$ in~\autoref{potencarte}  as a function of the field parameter
$\beta$. Dashed vertical lines and circles indicate the values of $\beta$ where bifurcations occur.}
\label{fi:energyCritical}
\end{figure}

\subsection{Bifurcations}
The bifurcations between the critical points are presented in~\autoref{fi:energyCritical} 
by the evolution of the equilibria energies as a function of the electric field parameter $\beta$. 
For the field-free case $\beta=0$, besides the
(isolated) equilibria $P_1$ and $P_5$, which are respectively the
{\sl head-tail} ground state (minimum) configurations and the {\sl head-head} and {\sl tail-tail} maxima, the
system presents the aforementioned two circles $D_{1,2}$ of degenerate equilibria,
see~\autoref{sec:equilibria}.
For $\beta>0$, the 
circles of equilibria blow-up, and from $D_1$ and $D_2$  emerge the two isolated equilibria 
$P_{3,4}$ and $P_{2,6}$, respectively.
In the interval $0< \beta < 1$, the study of the stability of the equilibria shows that
$P_1$ and $P_5$ are the absolute minima and maxima of the system, see~\autoref{fi:energyCritical},
 while $P_{2,3,4,6}$ are saddle points
of rank-one, rank-two, rank-three and rank-one, respectively.
As $\beta$ increases towards $\beta=1$, the equilibria $P_4$ and $P_5$ approach each other
see~\autoref{fi:energyCritical}, such that, at $\beta=1$, a first Pitchfork bifurcation takes place:
the two equilibria collide and  only the equilibrium $P_4$ survives afterwards, becoming
the equilibrium of maximal energy. This is the expected configuration of maximal energy
where both dipoles are located
along the negative $z_{1}$ and $z_{2}$ axes, \ie, oriented antiparallel to electric field direction.

In the interval $ 0<\beta<3$, the equilibria $P_{1,3,6}$ approach
each other (see~\autoref{fi:energyCritical}), and at $\beta=2$, equilibria $P_3$ and $P_6$ coincide. From this second
pitchfork bifurcation, only equilibrium $P_3$ survives, becoming
a saddle point of rank-one. Finally, at $\beta=3$, a third Pitchfork bifurcation between the
equilibria $P_1$ and $P_3$  occurs such that, for $\beta>3$, only $P_3$ survives becoming the
equilibrium of minimal energy.  
In this configuration of minimal energy,  both dipoles are  oriented along the field, \ie,  along the positive $z_{1,2}$ axes.
By further increasing $\beta$, $\beta > 3$, the dynamics  is gradually dominated by the 
interaction with the electric field, and the landscape of the potential energy surface
${\cal V}_1( \mathbf{l}_1, \mathbf{l}_2)$ resembles the one obtained by 
neglecting the dipole-dipole interaction.

{}

\end{document}